\documentclass{aa}  
\usepackage{graphicx}
\usepackage{txfonts}
\usepackage{gensymb}
\usepackage[]{hyperref}
\hypersetup{
    colorlinks=true,
    linkcolor=blue,
    filecolor=magenta,      
    urlcolor=cyan,
}

\begin{document} 
\title{ The X-ray emission of the Seyfert 2 galaxy MCG-01-24-12}
\titlerunning{X-ray spectrum of MCG-01-24-12}
\authorrunning{R. Middei, G. Matzeu et al.,}

\author{R. Middei
	\inst{1,2} \fnmsep\thanks{riccardo.middei@ssdc.asi.it},
	 G. A. Matzeu\inst{3,4}, S. Bianchi, \inst{5}, V. Braito \inst{6}, J. Reeves \inst{7},   A. De Rosa \inst{8}, M. Dadina \inst{9}, A. Marinucci \inst{10},\\ M. Perri \inst{1,2}, A. Zaino \inst{5} }
	 
\institute{
Space Science Data Center, SSDC, ASI, via del Politecnico snc, 00133 Roma, Italy
\and INAF - Osservatorio Astronomico di Roma, via Frascati 33, I-00078 Monteporzio Catone, Italy
\and Department of Physics and Astronomy - DIFA, University of Bologna, Via Gobetti 93/2 - 40129 Bologna, Italy
\and European Space Agency (ESA), European Space Astronomy Centre (ESAC), E-28691 Villanueva de la Ca$\tilde{n}$ada, Madrid, Spain
\and Dipartimento di Matematica e Fisica, Universit\`a degli Studi Roma Tre, via della Vasca Navale 84, I-00146 Roma, Italy
\and INAF-Osservatorio Astronomico di Brera, Via Bianchi 46, I-23807 Merate (LC), Italy
\and Department of Physics, Institute for Astrophysics and Computational Sciences, The Catholic University of America, Washington, DC 20064, USA
\and INAF/Istituto di Astrofisica e Planetologia Spaziali, via Fosso del Cavaliere, 00133 Roma, Italy.
\and INAF-Osservatorio di Astrofisica e Scienza dello Spazio, via Gobetti 93/3, 40129 Bologna BO.
\and ASI-Unit\`a di Ricerca Scientifica, Via del Politecnico snc, I-00133 Roma, Italy
}

% \abstract{}
% 5 {} token are mandatory
 
  \abstract
   {We present a detailed X-ray spectral analysis of the nearby Seyfert 2 galaxy MCG-01-24-12 based on a multi-epoch data set. Data have been taken with different X-ray satellites, namely  \textit{XMM-Newton}, \textit{NuSTAR}, \textit{Swift} and \textit{Chandra} and cover different time intervals, from years down to a few days. From 2006 to 2013 the source had a 2-10 keV flux of $\sim$1.5$\times$10$^{-11}$ erg cm$^{-2}$ s$^{-1}$, consistent with archival observations based on \textit{HEAO} and \textit{BeppoSAX} data, though a 2019 \textit{Chandra} snapshot caught the source in an extreme low flux state, a factor of $\sim$10 fainter than its historical one. Based on phenomenological and physically motivated models, we find the X-ray spectrum of MCG-01-24-12 to be best modelled by a power-law continuum emission with $\Gamma$=1.76$\pm$0.09 with a high energy cut-off at E$_{\rm c}=70^{+21}_{-14}$ keV that is absorbed by a fairly constant column density of N$_{\rm H}$=(6.3$\pm$0.5)$\times10^{22}$ cm$^{-2}$. These quantities allowed us to estimate the properties of the hot corona in MCG-01-24-12 for the cases of a spherical or slab-like hot Comptonising plasma to be kT$_{\rm e}$=27$^{+8}_{-4}$ keV, $\tau_{\rm e}$=5.5$\pm$1.3 and kT$_{\rm e}$=28$^{+7}_{-5}$ keV, $\tau$=3.2$\pm$0.8, respectively. 
   Finally, despite the short duration of the exposures, possible evidence of the presence of outflows is discussed.

   }

	\keywords{galaxies: active - galaxies: Seyfert - X-rays: galaxies - X-rays: individual: MCG-01-24-12}

   \maketitle
%
%-------------------------------------------------------------------
\section{Introduction}

\indent Seyfert 2 galaxies are a class of active galactic nuclei (AGNs) whose optical spectra lack broad emission lines. These emission lines are not observed because the Broad Line Region (BLR) is hidden by matter with column density in the range 10$^{22-24}$ cm$^{-2}$ and our line-of-sight passes through this obscuring circumnuclear medium, thought to be toroidal in structure. The so-called dusty torus absorbs and reprocesses (transmits and/or reflects) the X-ray nuclear continuum imprinting some characteristic features on the emerging spectrum. Although this obscuring structure is predicted to be ubiquitous in AGN by the unification model \citep{Antonucci1993}, there are still uncertainties on its exact location, composition and overall geometry. In recent years both short- and long- term variability of the obscurer column density (N$_{\rm H}$) have been observed in nearby AGNs such as  NGC 1365 \citep[e.g.][]{Risaliti05b,Rivers2015}, NGC 4388 \citep[e.g.][]{Elvis2004}, NGC 4151 \citep[][]{Puccetti2007}, and NGC 7582 \citep[e.g.][]{Piconcelli2007,Bianchi2009,Rivers2015,Braito2017} and changes in the column density have also been measured in heavily obscured AGN, i.e. NGC 1068 \citep[e.g.][]{Zaino2020a}. As a result, the standard picture of a smooth `doughnut' shaped torus was ruled out in favour of a more inhomogeneous structure comprised by a distribution of a large number of individual clumps \citep[e.g.,][]{Risaliti02,Markowitz14}.\\
The high piercing power of X-rays, unless the obscuring matter has a column density larger than the inverse of the Thomson cross section (N$_{\rm H}>\sigma_{\rm Th.}^{-1}\sim1.5\times10^{24}$ cm$^{-2}$), can allow for the direct observation of the central regions where the primary X-rays originate \citep[e.g.][]{Nandra1994,Turner1997a,Guainazzi2005,Awaki2006}. X-ray photons are produced by inverse-Compton scattering disc optical-UV photons from thermal electrons, the so-called hot corona \citep[details in][]{haar91,haar93}. Both variability and microlensing arguments \citep[][]{Chartas09, Morgan12,DeMa13,Kara16} agree with this hot plasma being compact and likely lying close to the SMBH. As expected on a theoretical basis, the physical quantities of the X-ray emitting region (its opacity and temperature) characterise the emerging spectrum in terms of photon index ($\Gamma$) and high energy cut-off \citep[E$_{\rm c}$, e.g.][]{Rybi79,Ghisellini13}. The relation between physical and phenomenological quantities has been the object of different studies \citep[e.g.][]{Belo99,Petr00,Petr01,Middei2019b}. 
\textit{NuSTAR} played a fundamental role in such a framework; due to its unrivalled capability of focusing X-rays up to about 80 keV, it allowed for an increasingly number of high energy cut-off measurements \citep[e.g.][]{Fabi15,Fabi17,Tortosa18}, hence estimates of the coronal temperature (kT$_{\rm e}$) and optical depth ($\tau_{\rm e}$).\\ 
\indent The primary X-ray continuum can be reflected off the black hole surroundings and this emission may leave two major signatures on the emerging spectrum: a Fe K$\alpha$ line at 6.4 keV due to fluorescence and a bump of counts at $\sim$30 keV \citep[e.g.][]{George91,Matt91}, the so-called Compton hump.\\
\indent Due to their variable emission, multi epoch data sets are particularly suitable for studying AGNs spectral properties. In this context, we report on the  analysis of MCG-01-24-12, a bright spiral galaxy at redshift $z$=0.0196 \citep[e.g.][]{Koss2011} hosting a Seyfert 2 nucleus \citep{deGrijp1992}. This AGN, firstly studied by \citet{Piccinotti82} in the 2-10 keV band, has been subsequently analysed by \citet{Malizia2002} that used \textit{Beppo-SAX} data to discuss its X-ray spectral properties. In particular, the source 2-10 keV band was characterised by a narrow Fe K$\alpha$ emission line and an absorption feature at about 8 keV, while softer X-rays were absorbed by an obscurer with N$_{\rm H}\sim7\times10^{22}$ cm$^{-2}$. In the hard X-band (20-100 keV) they measured a flux of $\sim$4$\times$ 10$^{-11}$ erg cm$^{-2}$ s$^{-1}$. Later, MCG-01-24-012 has been reported by \cite{Ricci2017,Ricci18} who, using \textit{Swift} data, measured its flux to be F$_{\rm 14-195~keV}$=4.1$\times$10$^{-11}$ erg cm$^{-2}$ s$^{-1}$. Finally, MCG-01-24-12 black hole mass has been estimated to be M$_{\rm BH}$=1.5$^{+1.1}_{-0.6}\times10^{7}$ M$_\odot$ \citep[][]{LaFranca2015}.

\section{Archival data}
\indent MCG-01-24-12 was observed several times in the X-ray band. In Table~\ref{log} we report the log of the observations considered in this work.
	\begin{table}
	\centering
	\setlength{\tabcolsep}{0.7pt}
	\caption{\small{The observation log of the presented data set is reported. For four pointings out of six, the \textit{NuSTAR} and \textit{Swift} observatories have simultaneously observed MCG-01-24-12.}\label{log}}
	\begin{tabular}{c c c c c c c}
		\hline
		\\
		Satellite &Detector&Obs. ID& Obs.&Net exposure & Start-date & \\
		  & & & &ks & yyyy-mm-dd & \\
		\\
		\textit{XMM}&\textit{pn/MOS}&0307000501 &&8.3/15.4&2006-04-25&\\
		\\
		\textit{NuSTAR}&\textit{FPMA/B} &60061091002	 &1&12.3&2013-04-03&\\
		\textit{Swift}&\textit{XRT} 	 &00080415001&&7.5&\\
		\\
		\textit{NuSTAR}&\textit{FPMA/B} &60061091004	 &2&9.3&2013-04-10&\\
		\textit{Swift}&\textit{XRT}  	 &00080415002&&1.9&\\
		\\
		\textit{NuSTAR}&\textit{FPMA/B} &60061091006&	 3&12.1&2013-04-18&\\
		\textit{Swift}& \textit{XRT} &00080415003&&2.9&\\
		\\
		\textit{NuSTAR}&\textit{FPMA/B} &60061091008&	 4&14.0&2013-05-05&\\
        \\
        \textit{NuSTAR}&\textit{FPMA/B} &60061091010&	 5&15.3&2013-05-12&\\
        \textit{Swift}& \textit{XRT} 	&00080415005&&1.8&\\
        \\
        \textit{NuSTAR}&\textit{FPMA/B} &60061091012&	6 &12.2&2013-05-22&\\
         \\
        \textit{Chandra}&\textit{ACIS-S} &703907& &9.1&2019-06-27&\\
        \hline
		\hline
	\end{tabular}
\end{table}

\begin{figure}
	\includegraphics[width=0.49\textwidth]{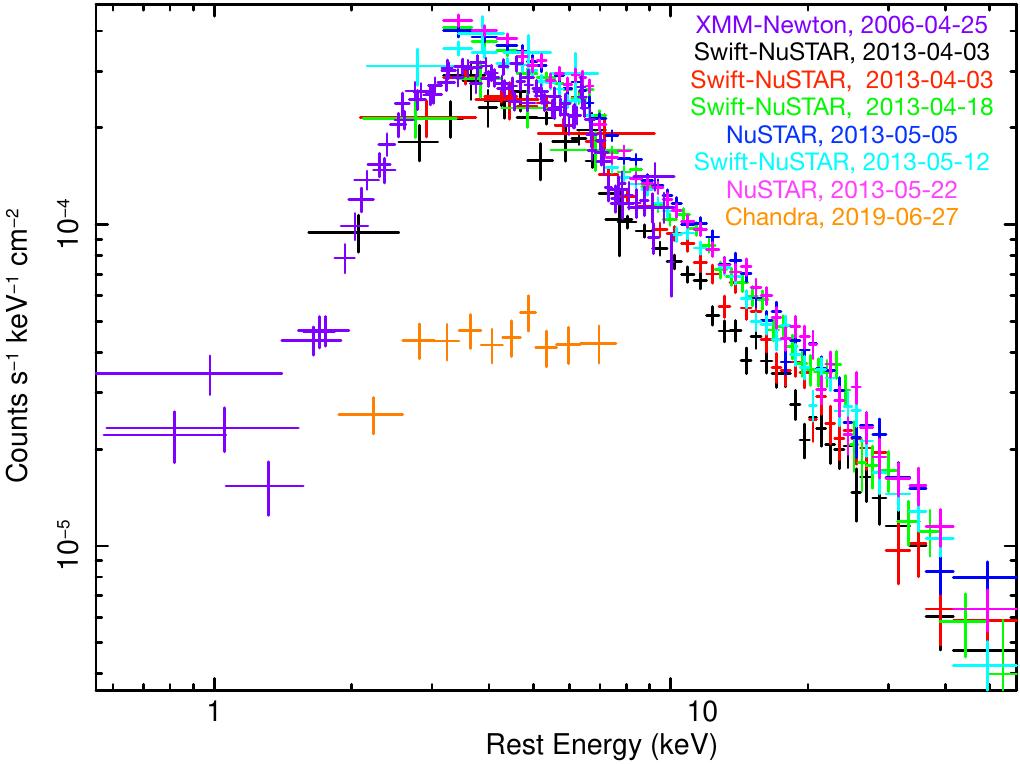}
	\caption{\small{MCG-01-24-12 spectra as observed by the different observatories as folded by a power-law  with $\Gamma$=2  and  unitary normalisation.\label{all_data}}}
\end{figure}
\begin{figure*}
	\includegraphics[width=0.99\textwidth]{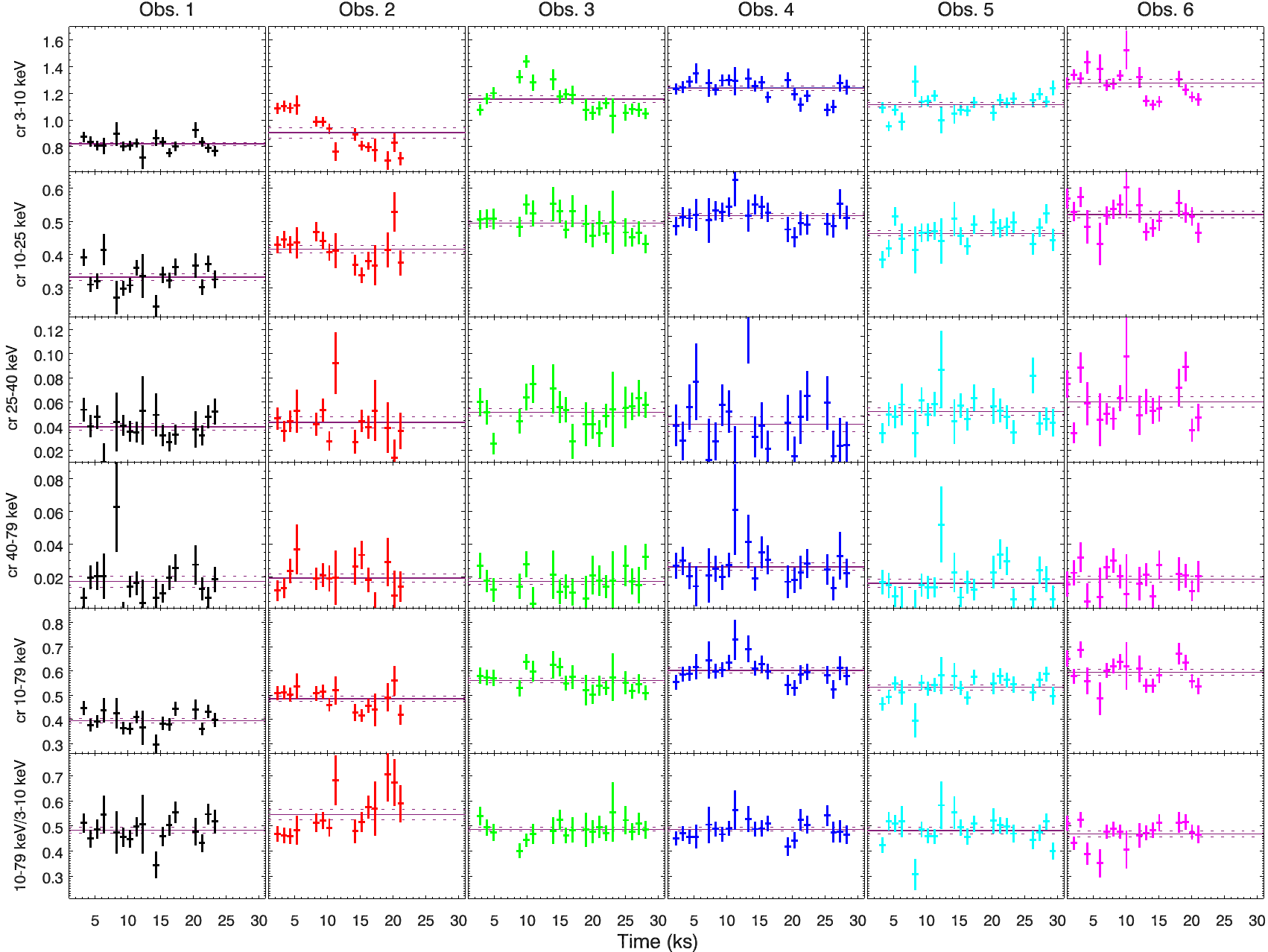}
	\caption{\small{Background subtracted \textit{NuSTAR} light curves of MCG-01-24-12 calculated with a temporal bin of 1000 sec. Light curves account for co-added module A and B and the various energy bands are labelled on the y-axis. For each observation a specific colour is used and  we adopted such a colour code throughout the whole paper. Orchid horizontal straight lines are used to quantify the average count rate within each observation for the different bands, while dashed ones account for 1$\sigma$ uncertainties around the mean}.\label{lc}}
\end{figure*}
 The standard \textit{XMM-Newton} Science Analysis System ($SAS$, Version 18.0.0) has been used to obtain the scientific products for the different instruments on-board the observatory, namely \textit{pn} \citep{Stru01} and the \textit{MOS} cameras \citep{Turn01}. To select the extraction radius and to screen for high background time intervals we used an iterative process that maximises the Signal-to-Noise Ratio (S/N) ratio \citep[ see details in][]{Pico04}. For the \textit{pn} data, we used a 21 arcsec radius circular region to extract the source spectrum and the background was computed with a circular area of 50 arcsec radius close to the source. The spectrum was then binned to have at least 30 counts for each bin, and not to oversample the instrumental energy resolution by a factor larger than 3. Radii of 21 and 22 arcsec were adopted for \textit{MOS1} and \textit{MOS2}, respectively, to extract the source spectrum, while we obtained the background using a circular area with 40 arcsec radius. The same binning strategy used for \textit{pn} data has been applied to \textit{MOS} spectra. We notice that these \textit{XMM-Newton} data are not affected by significant pile-up this being in agreement with the \textit{SAS} standard task \textit{epatplot}. Moreover, we checked for the eventual $Cu$ emission possibly affecting the \textit{pn} spectrum and no evidence of it was found.\\
\indent To calibrate and clean raw \textit{NuSTAR} data we used the NuSTAR Data Analysis Software (\textit{NuSTARDAS}, Perri et al., 2013\footnote{\url{https://heasarc.gsfc.nasa.gov/docs/nustar/analysis/nustar\_swguide.pdf}}) package (v. 1.8.0). Level 2 cleaned products were obtained with the standard \textit{nupipeline} task and the scientific products were computed thanks to the \textit{nuproducts} pipeline and using the calibration database "20191219". A circular region with a radius of 40 arcsec was used to extract the source spectrum. The background has been calculated using the same circular region but centered in a blank area nearby the source. The \textit{FPMA/B} spectra have been binned in order not to oversample the instrumental resolution by a factor larger than 2.5 and to achieve a S/N larger than 3 in each spectral channel.\\
\indent \textit{Swift-XRT} data were taken in photon counting mode and we derived the scientific products using the facilities provided by the Space Science Data Center, (SSDC, \url{https://www.ssdc.asi.it/}) of the Italian Space Agency (ASI).
The source spectrum was extracted with a circular region of $\sim$ 60 arcsec and we used a concentric annulus for the background. Then spectra have been binned in order to have at least 20 counts in each bin. Due to their short exposures, we do not show the \textit{XRT} light curves.\\
\indent An approximately 10 ks long exposure of MCG-01-24-12 was carried out by \textit{Chandra} on the 27th June 2019 with the Advanced CCD Imaging spectrometer (ACIS-S; \citealt{Garmire2003}). The data was reduced by adopting the Chandra Interactive Analysis of Observation software (\textit{CIAO} v. 4.12 \citealt{Fruscione06}) and the latest Chandra Calibration Data Base (\textit{CALDB} version 4.9.2.1). The source and background spectra were extracted using a circular region of $2.5''$ and $4.0''$ radius respectively. Furthermore the resulting spectrum was re-binned by a minimum of 20 counts per energy bin and with a total net count of 610  for a net exposure time of $9133\,s$.\\
\indent In all the fits of simultaneous \textit{Swift-NuSTAR} observations, the inter-calibration between the different \textit{NuSTAR} modules and the \textit{Swift}'s X-ray telescope is taken into account by a cross-normalisation constant. The  \textit{FPMA/B} modules were always found consistent within a 3\% with the exception of observation 1 in which they agree within a 30\%. In particular, the \textit{FPMB} spectrum has about 1000 counts more than \textit{FPMA/B}. In this latter module, the source was found to lie between the detector chips, this explains the decreased number of photon counts. However, by fitting individually the \textit{FPMA/B} spectra with an absorbed power-law the returned photon indices were consistent within the errors, hence we decided to include data from both modules in the forthcoming analysis. In all but one pointing \textit{Swift/XRT} and \textit{NuSTAR} are consistent within $\lesssim$10\%, in accordance with \cite{Madsen2015}. On the other hand, for Obs. 4 we obtained const=1.5$\pm$0.3 as \textit{NuSTAR} caught the source in a higher flux state than the shorter \textit{Swift} snapshot. For a quick look comparison of the data, we show in Fig.~\ref{all_data} all the data. Finally, \textit{MOS1} and \textit{MOS2} are consistent with \textit{pn} data within 3\%. \\
\indent We preliminarily computed the MCG-01-24-12 light curves for the \textit{NuSTAR} data. In Fig. ~\ref{lc}, we report the corresponding time series in various bands as labelled on the y-axis while in the last row we show the ratios between the 3-10 and 10-79 keV bands. X-ray variability is typical of AGNs and it has been measured down to ks timescales up to decades \citep[e.g.][]{Green1993,Uttley2002,Vagnetti2011,Ponti2012,McHardy2017,Vagnetti2016, Middei2017,Paolillo2017}. Regarding MCG-01-24-12, Observation 2 varies by a factor of 50\% in the 3-10 keV energy band over a few ks while the other exposures have a more constant behaviour in the same energy band. On the other hand, variations are witnessed when comparing light curves at different observing epochs. The maximum amplitude change, about a factor of 2, occurred in the 3-10 keV energy band over a timescale of about 1 month (between observations 1 and 6). The ratios between the 3-10 keV and the 10-79 keV bands have a rather constant behaviour, with the exception of Obs. 2 in which the source hardness increased as the flux decreased. 
However, the short exposure does not allow to perform a time-resolved spectral analysis of Obs. 2, hence we used the time-averaged spectra to improve the fit statistics of all the observations. All errors reported in the plots account for 1$\sigma$ uncertainty, while  errors  in  text  and  tables  are  quoted  at  90\%  confidence level.
\section{Spectral Analysis}
We used \textit{XSPEC} \citep{Arna96} to fit the data with the Galactic column density kept frozen to the value N$_{\rm H}$=2.79$\times$10$^{20}$ cm$^{-2}$ \citep{HI4PI}. Moreover, the standard cosmology \textit{$\Lambda$CDM} (H$_0$=70 km s$^{-1}$ Mpc$^{-1}$, $\Omega_m$=0.27, $\Omega_\lambda$=0.73) is adopted throughout the analysis.

\subsection{The Epic cameras view: 0.3-10 keV band}

\begin{figure}
	\includegraphics[width=0.49\textwidth]{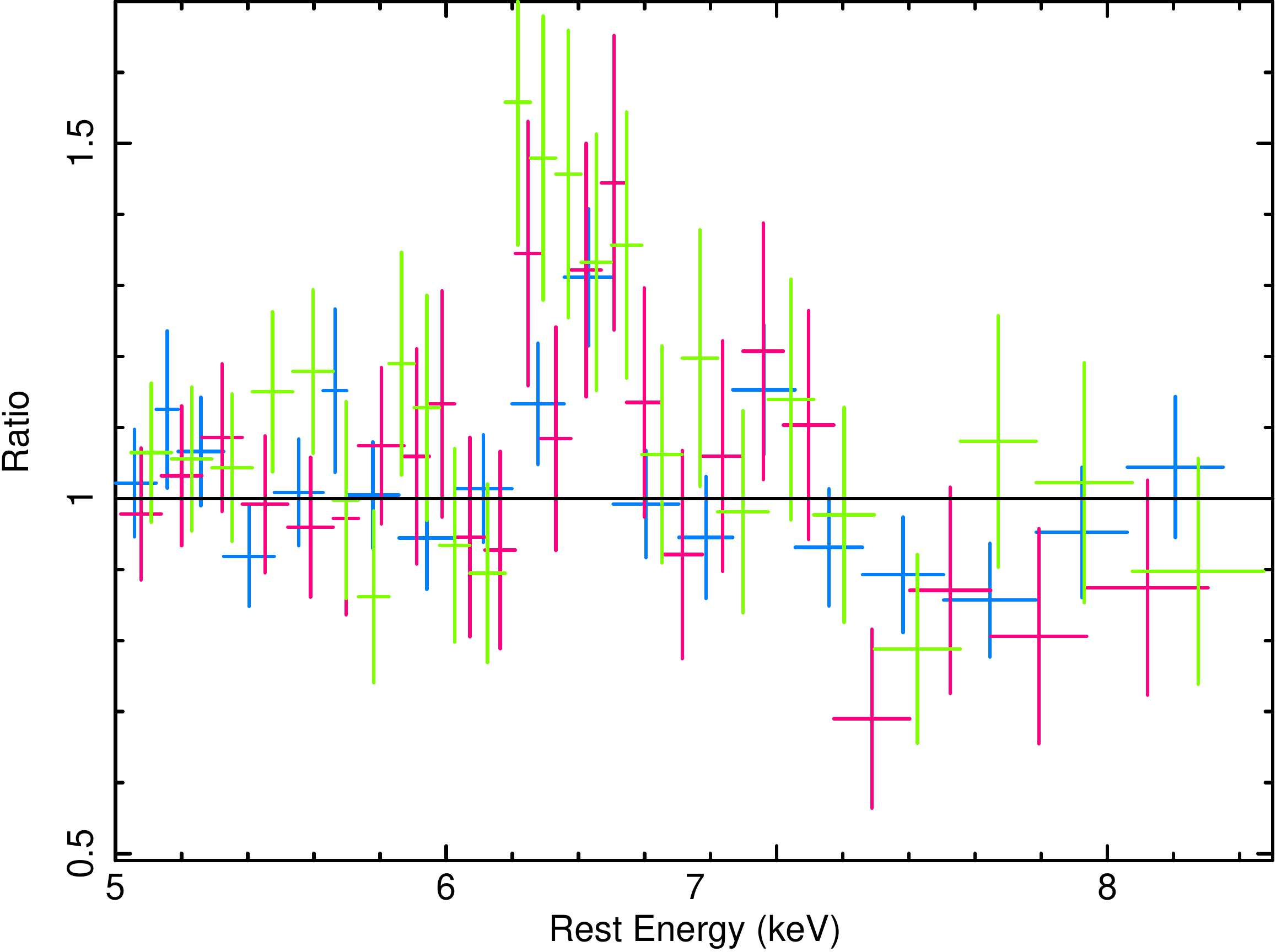}
	\includegraphics[width=0.5\textwidth]{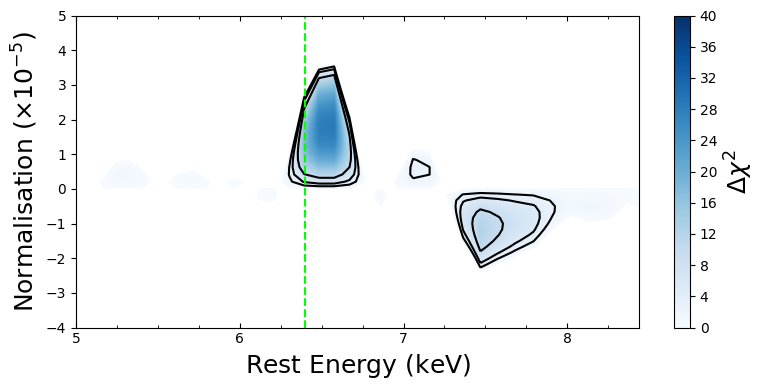}
	\caption{\small{ Top panel: Ratio of the Epic spectra in the 5-8.5 keV energy range with respect to an absorbed power-law. All the instruments (\textit{pn} in blue, \textit{MOS1} in magenta and \textit{MOS2} in green) detected the Fe K$\alpha$ and found a trough at about 7.5 keV.  Bottom panel: Blind line scan result between the normalisation and
	line energy, adopted on the \textit{pn-MOS} spectra, where a Gaussian is left free to vary in the 5-8.5\,keV energy range. The colour bar on the right indicates the significance of the lines for 2 degree of freedom and the solid black contours correspond to $68\%$, $90\%$ and $99\%$.
	\label{cont_pn_mos}}}
\end{figure}

\begin{figure}
	\includegraphics[width=0.49\textwidth]{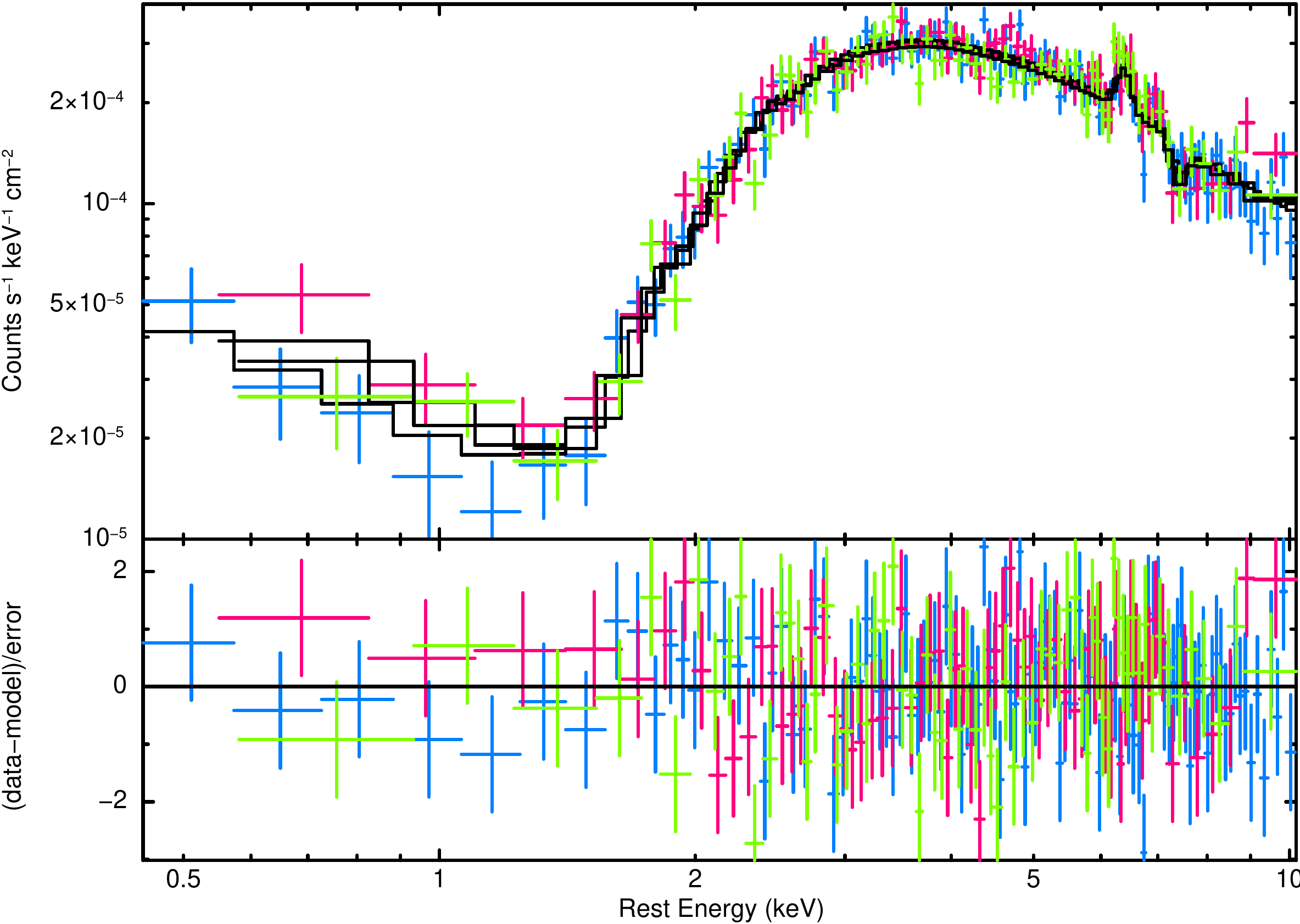}
	\caption{\small{Best-fit model to the EPIC-pn\&MOS with a corresponding statistics of $\chi^2$=243 for 237 d.o.f.~.\label{pn_mosbf}}}
\end{figure}

We started studying the 0.3-10 keV \textit{EPIC} data using a simple model (\textit{const$\times$tbabs$\times$ztbabs$\times$power-law}, in \textit{XSPEC} notation)  whose components account for the inter-calibration between the cameras, the Galactic absorption, the MCG-01-24-12 intrinsic absorption and its primary continuum emission. Such a model leaves prominent residuals in the soft X-ray band and the fit is unacceptable on statistical grounds ($\chi^2$=355 for 243 d.o.f.). The excess in the soft X-rays may be due to a fraction of the coronal primary emission that is scattered/reflected possibly by distant material.
Such a behaviour in obscured AGNs is quite typical. This energy band is generally dominated by emission lines from a photonionised gas coinciding with the Narrow Line Region \citep[NLR, e.g.][]{Bianchi2006,Guainazzi2007} though the lack of high statistics or the low-resolution of X-ray spectra makes it possible to model such a component with a simple power-law \citep[e.g.][]{Awaki1991,Turner1997a,Turner1997b}.
Since our spectra do not show any of these features, we model the scattered component in the \textit{EPIC} data adding an additional power-law.
A test fit has shown that the photon indices of the power-laws accounting for the primary X-ray continuum and the scattered emission to be compatible within the errors, hence, in the following fits, we tied these parameters. The normalisation of the scattered component has been fitted and found to be a few percent of the primary continuum \citep[e.g.][]{Bianchi2007}. The addition of this new component is beneficial in terms of statistics and the fit improved by $\Delta\chi^2/\Delta$d.o.f.=-83/-1. \\
\indent Residuals are still visible in the Fe K$\alpha$ band and are reported in the top panel of Fig.~\ref{cont_pn_mos} and further supported by the blind lines scan shown in the bottom panel of the same figure. The blind line scan was carried out by using an absorbed power-law and a Gaussian model with a fixed line width of 1 eV (i.e. much lower than the CCD resolution)  with line energy allowed to vary in the range $5-10\,\rm keV$ and normalisation from $-6\times10^{-5}$ to $10^{-4}\,\rm ph\,keV^{-1}\,cm^{-1}\,s^{-1}$ with 50 steps.  This test suggests the presence of a strong Fe K$\alpha$ emission line at the rest-frame energy of $E=6.39\pm0.05~\rm keV$ and an absorption trough at $\sim7.4\pm0.1~\rm keV$ at $\sim5\sigma$ and $\sim3\sigma$ confidence levels respectively. We accounted for these features by adding two Gaussian components, one for reproducing the Fe K$\alpha$ feature, the other the trough above $\sim$7 keV. For both these components we fitted the line's energy centroid, its width and normalisation. Since we only get an upper limit for the width of the absorption component $\sigma_{\rm Fe~K}$<160 eV, we fixed this parameter to this value. The addition of the Gaussian emission improved the overall statistic by $\Delta \chi^2$=18 for 3 d.o.f. less, while a $\Delta\chi^2/\Delta$d.o.f.=-10/-2 corresponds to the line in absorption. The described procedure led to a best-fit of $\chi^2$=243 for 237 d.o.f. for which we report best-fit parameters in Table \ref{xmmfit}.
The \textit{EPIC} spectra are therefore well described by a power-law continuum  ($\Gamma=1.68\pm0.04$) absorbed by an obscurer with column density of N$_{\rm H}=6.4\pm0.3\times10^{22} \rm cm^{-2}$. The continuum is accompanied by a moderately broad ($\sigma=80\pm60$ eV) neutral Fe emission line and a weak absorbing signature.
Finally, the soft X-ray band is dominated by a fraction of the primary continuum scattered, see Fig.~\ref{pn_mosbf}.

\subsection{Broad-band \textit{Swift/NuSTAR} view: I the phenomenological model}

\begin{table}
	\centering
	\setlength{\tabcolsep}{1.5pt}
	\caption{\small{The \textit{XMM-Newton} best-fit values ($\chi^2$=243 for 237 d.o.f.) as derived with  the simple phenomenological model reported in Sect. 3.1.~.The crux desperationis $\dagger$ is used to identify  a frozen parameter.}\label{xmmfit}}
	\begin{tabular}{c c c}
		\hline
		\\
		Parameter &Best-fit value&Units \\
		N$_{\rm H}$&6.4$\pm$0.3&$\times$10$^{22}$ cm$^{-2}$\\
		$\Gamma$&1.68$\pm$0.04&\\
		N$_{\rm primary}$&4.7$\pm$0.3 & $\times$10$^{-3}$ ph.\,keV$^{-1}$\,$\rm cm^{2}$\,s$^{-1}$\\
		E$_{\rm Fe~K\alpha}$&6.39$\pm$0.05& keV\\
		$\sigma_{\rm Fe~K\alpha}$&80$\pm$60& eV\\
		EW$_{\rm Fe~K\alpha}$&95$\pm$30& eV\\
		N$_{\rm Fe~K\alpha}$&2.5$\pm$0.7  &$\times$10$^{-5}$ ph.\,cm$^{-2}$\,s$^{-1}$\\
		E$_{\rm Fe~K}$&7.4$\pm$0.1& keV\\
		$\sigma_{\rm Fe}\dagger$&160& eV\\
		EW$_{\rm Fe~K}$&-70$\pm$35& eV\\
		N$_{\rm Fe~K}$&-1.3$\pm$0.6 &$\times$10$^{-5}$ ph.\,cm$^{-2}$\,s$^{-1}$\\
		N$_{\rm scattered}$&1.9$\pm$0.4& $\times$10$^{-5}$ ph. keV$^{-1}$ cm$^{-2}$ s$^{-1}$\\
		F$_{\rm 2-10~keV} $&1.3$\pm$0.1 &$\times$10$^{-11}$ erg cm$^{-2}$ s$^{-1}$ \\
		
	\end{tabular}
\end{table}

\begin{figure}
	\includegraphics[width=0.49\textwidth]{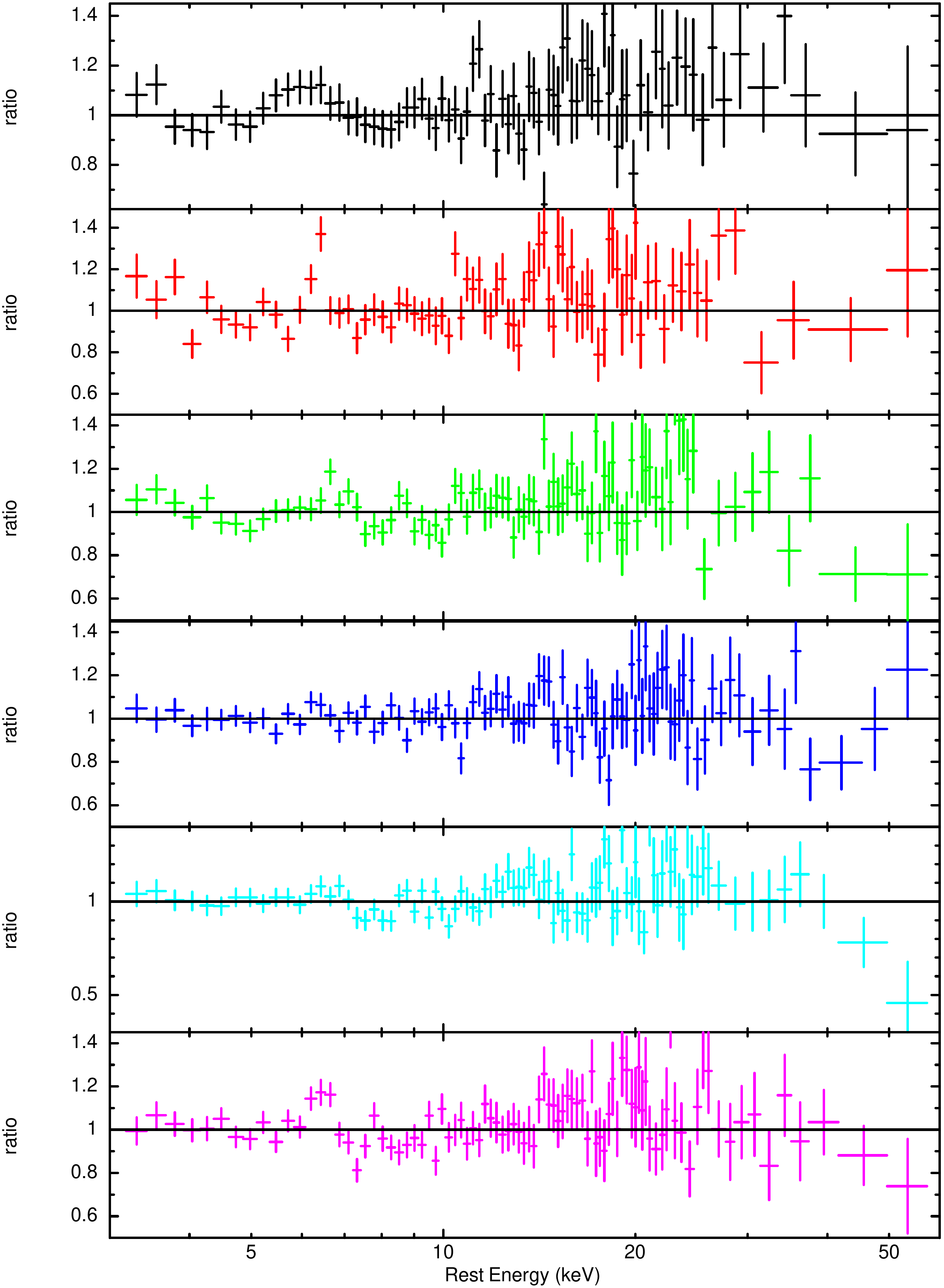}
	\caption{\small{Ratios between the \textit{NuSTAR} \textit{FPMA/B} spectra. The underlying model accounts for an absorbed power-la \textit{const$\times$tbabs$\times$ztbabs$\times$po} for each observation.\label{abspo}}}
\end{figure}

\begin{figure*}
 \centering
	\includegraphics[width=0.99\textwidth]{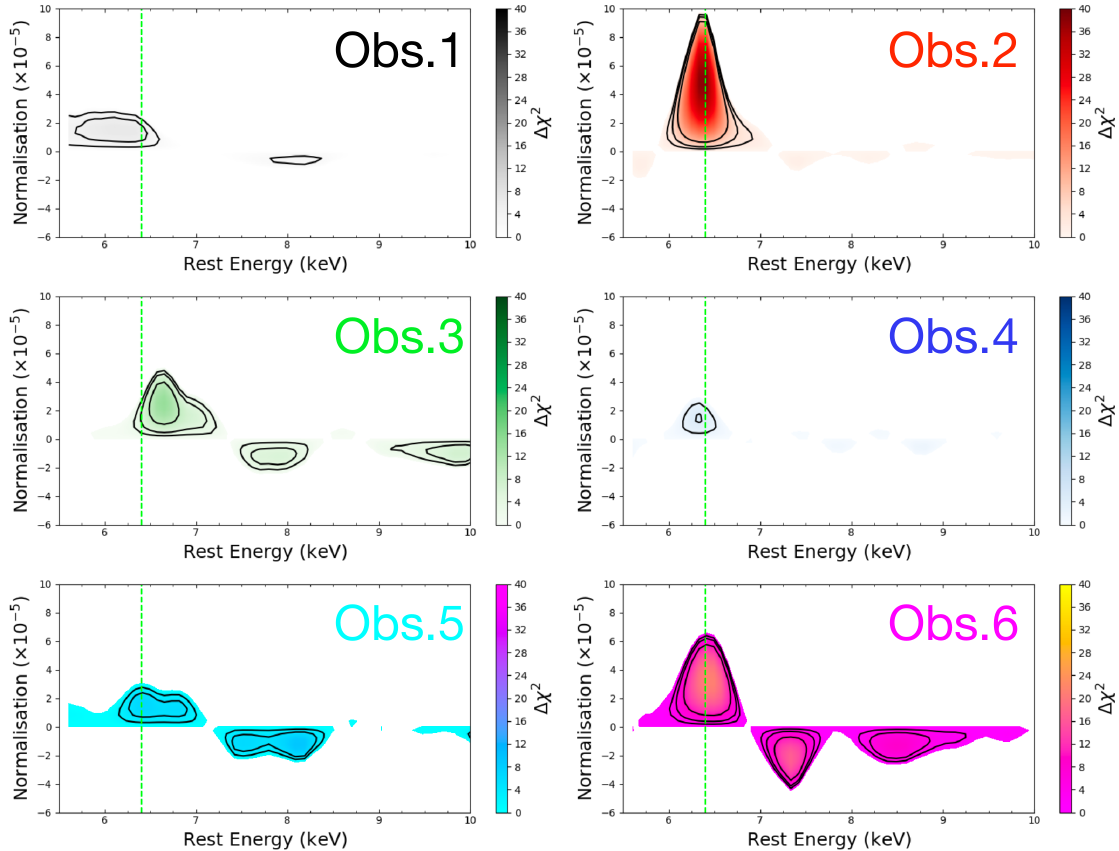}	
	\caption{\small{Result of a blind line scan performed to all six \textit{NuSTAR} observations plotted between the $5.5-10\,\rm keV$ band. The colour bar on the right of each panel indicates the significance of the lines for 2 degree of freedom and the solid black contours correspond to $68\%$, $90\%$ and $99\%$. The lime-green vertical dashed line indicates the position of the rest frame energy of the Fe\,K$\alpha$ emission line at $E=6.4\,\rm keV$\label{scannu}}}
\end{figure*}

We started analysing the \textit{NuSTAR} data by fitting each observation separately with an absorbed power-law and the corresponding residuals are shown in Fig~\ref{abspo}.
To better quantify the possible variability of the Fe K$\alpha$ and to further investigate for the presence of any additional features, we performed a blind line scan over the spectra. The line scan procedure was the same described for the \textit{XMM-Newton} data. The resulting contour plots are reported in Fig~\ref{scannu}. The Fe K$\alpha$ line appears stronger in Observations 2 and 6 while it is only marginally detected in all the remaining pointings. Interestingly, in Obs. 6, the energy spectrum suggests the presence of an absorbing signature above $\sim$7 keV. A fit of such a feature with a narrow Gaussian component returns E=7.35$\pm$0.10 keV, N=(1.8$\pm$0.8)$\times$10$^{-5}$  ph. cm$^{-2}$ s$^{-1}$ and EW=-80$\pm$35 eV, with an improvement of the fit statistics of $\Delta\chi^2$/d.o.f.=-14/-2. We then considered also the \textit{Swift} data and fitted the  \textit{Swift-NuSTAR} observations using a Gaussian line to account for the Fe K$\alpha$ emission plus an absorbed cut-off power-law for reproducing the continuum, written in \textit{XSPEC} syntax as \textit{tbabs$\times$ztbabs$\times$(cutoffpl+zgauss)}.
The high energy cut-off for the primary emission is included in order to model the counts drop observed above $\sim$40 keV in Fig.~\ref{abspo} for Obs. 3, 5 and 6.
 The current data, with Obs. 6  being a possible exception, does not allow for an appropriate characterisation of the troughts in the spectra, thus we did not account for them in the modelling.
The fitting procedure was performed allowing the photon index, the high energy cut-off and the continuum normalisation to vary in each observation. Then we calculated the emission line normalisation, while the energy centroid was set to 6.4 keV. We assumed the line to have a narrow profile with $\sigma$ fixed to 1eV. After a preliminary fit showing the obscurer column density to be consistent within the uncertainties in all the pointings, we tied the \textit{ztbabs} between the different observations. The \textit{Swift-XRT} data have too poor statistics to constrain any scattered emission (see Fig.~\ref{all_data}) and, for this reason, we do not include an additional power-law accounting for such a component. This fit leads to a fit characterised by $\chi^2$/d.o.f.=1309/1256 and in Table~\ref{pippo} we quote the corresponding best-fit parameters.

\begin{table*}
	\centering
	\setlength{\tabcolsep}{5.pt}
	\caption{\small{\textit{Swift-NuSTAR} best-fit parameters derived using an absorbed power-law plus a Gaussian component accounting for the Fe K$\alpha$ emission line (  \textit{tbabs$\times$ztbabs$\times$(zgauss+power-law)} in \textit{XSPEC}) corresponding to $\Delta\chi^2$/d.o.f.=1309/1256. The $\dagger$ labels those parameters that have been fitted but tied between the observations.}}\label{pippo}
	\begin{tabular}{c c c c c c c c}
		\hline
		\\
		Parameter & Obs. 1 &Obs. 2 & Obs. 3& Obs. 4& Obs. 5& Obs. 6&Units\\
		N$_{\rm H}\dagger$&6.6$\pm$0.4 & & & & &&$\times$10$^{22}$ cm$^{-2}$ \\
		$\Gamma$&1.81$\pm$0.04&1.73$\pm$0.04&1.79$\pm$0.03 &1.77$\pm$0.04&1.77$\pm$0.04&1.77$\pm$0.03&\\
		N$_{\rm primary}$&4.6$\pm$0.5&4.7$\pm$0.6&5.2$\pm$0.6 &8.0$\pm$0.6 &7.4$\pm$0.9&7.9$\pm$0.5&$\times$10$^{-3}$~ph. keV$^{-1}$\,$\rm cm^{2}$\,s$^{-1}$\\
		E$_{\rm Fe K\alpha}\dagger$&6.4 &   &  & &  &&keV \\
		EW$_{\rm Fe~K\alpha}$&85$\pm$50&210$\pm$60&43$\pm$38&45$\pm$35&55$\pm$35&105$\pm$40&eV\\
		N$_{\rm Fe~K\alpha}$&1.4$\pm$0.8&4.1$\pm$1.2&0.85$\pm$0.76&1.4$\pm$1.1&1.6$\pm$1.0&3.3$\pm$1.2&$\times$10$^{-5}$~ph. cm$^{-2}$ s$^{-1}$  \\
		F$_{\rm 3-10~keV} $& 1.23$\pm$0.08& 1.41$\pm$0.18&1.44$\pm$0.16& 2.23$\pm$0.10&2.13 $\pm$0.23&2.29$\pm$0.11&$\times$10$^{-11}$ erg cm$^{-2}$ s$^{-1}$  \\
	\end{tabular}
\end{table*}
%so ∆EW=(∆norm/norm)*EW

The current phenomenological model is consistent with a power-law with constant shape that is absorbed by an average column density N$_{\rm H}$=(6.6$\pm$0.4)$\times$10$^{22}$ cm$^{-2}$. Moreover, the Fe K$\alpha$ emission line, assumed to have a narrow intrinsic profile, seemed to be variable, at least on Observations 2 and 6. To further assess the actual variability of such a component, we assumed the emission line flux to be the same over the whole campaign. In other words, starting from the phenomenological best-fit, we fitted the data tying the Fe K$\alpha$ normalisation between the observations. The obtained fit has a worse statistics ($\Delta \chi^2$/$\Delta$d.o.f.=+23/+5), thus further supporting the line to be variable with significance larger than 95\%. 

\subsection{Broad-band Swift/NuStar view: II the reflection signature with \textit{Xillver}}
\label{subsec:The XRT-FPMA/B 0.5-79 band I: xillver}
As a subsequent step, we studied the \textit{XRT-FPMA/B} spectra using \textit{xillver} \citep[e.g.][]{Garc14a,Daus16} a model that self-consistently calculates the underlying nuclear emission (a cut-off power-law)  and the reflected component from the illuminated accretion disc. Then, we left free to vary and compute in each observation the photon index, the reflection parameter $R$ and the model normalisation. We considered the reflecting matter close to be neutral, we fixed the ionisation parameter $\log\xi$ to a value of 0 and we calculated the Fe abundance (A$_{\rm Fe}$) tying its value between the different observations. Though we know the  \textit{ztbabs} component does not vary significantly during the monitoring, we fitted this component in all the observations. Such steps led to the best-fit parameters in Table~\ref{xillver} and characterised by a statistic of $\chi^2$/d.o.f.=1252/ 1258.
The overall fit is consistent with a primary emission continuum absorbed by an average column density of N$_{\rm H}$=(6.3$\pm0.5$)$\times$10$^{22}$ cm$^{-2}$  and the reflecting matter arising from a cold disc region with a Solar metal abundance A$_{\rm Fe}$=1.3$\pm$0.6.
The photon index of the primary continuum varies in the range 1.68-1.93, though the obtained best-fit values are  consistent within the uncertainties. In similar fashion, marginal variability is observed for the high energy cut-off and the reflection parameter, and, as expected from the light curves in Fig. \ref{lc}, the normalisation of \textit{xillver} is found to vary. We report best-fit parameters in Table~\ref{xillver} and contours plot referring to the photon index, the reflection fraction and the high energy cut-off are shown in Fig~\ref{contnu}.\\
\indent We further tested the data using the same model but tying the photon index, the reflection fraction, the high energy cut-off and the column density of the obscurer between the exposures. The fit returns a statistic of $\chi^2$ 1300 for 1278 d.o.f., not far from the previous one, and further supports a weak variability of the parameters. \\
\begin{figure}
	\includegraphics[width=0.49\textwidth]{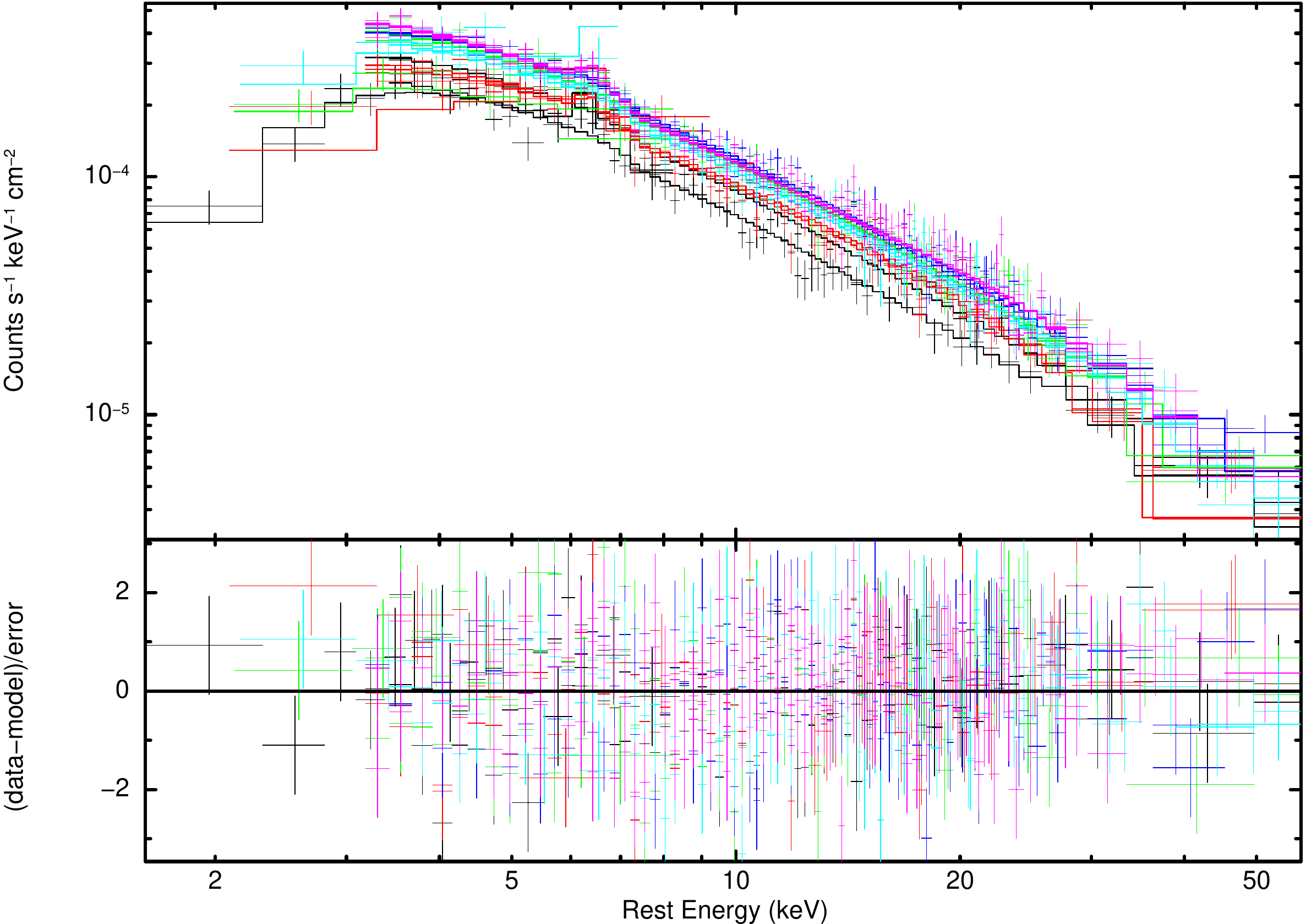}
	\caption{\small{Best-fit ($\chi^2$/d.o.f.=1252/1257) to the simultaneous \textit{Swift-NuSTAR} data obtained using \textit{xillver} \label{xill}.} }
\end{figure}
 \begin{table*}
	\centering
	\setlength{\tabcolsep}{2.5pt}
	\caption{\small{\textit{Swift-NuSTAR} best-fit values derived using \textit{tbabs$\times$ztbabs$\times$xillver} in \textit{XSPEC} notation and to which corresponds a statistic $\chi^2$/d.o.f.=1252/1257.}\label{xillver}}
	\begin{tabular}{c c c c c c c c c }
		\hline
		\\
		Parameter&Obs. 1 &Obs. 2 &Obs. 3 &Obs. 4&Obs. 5&Obs. 6 &all tied&Units\\
	    N$_{\rm H}$&7.2$\pm$1.0&7.2$\pm$1.9&6.1$\pm$0.9&6.2$\pm$1.1&6.2$\pm$0.9&6.1$\pm$	1.0&6.3$\pm$0.5&$\times10^{22}$ cm$^{-2}$\\
		$\Gamma$&1.93$\pm$0.12&1.68$\pm$0.15&1.76$\pm$0.11&1.71$\pm$0.09&1.78$\pm$0.09&1.89$\pm$0.10&1.76$\pm$0.09&\\
		 $R$&0.50$\pm$0.25&0.96$\pm$0.40&0.50$\pm$0.20&0.20$\pm$0.15&0.50$\pm$0.20&0.70$\pm$0.25&0.40$\pm$0.15&\\
		E$_{\rm c}$&>75&40$^{+30}_{-10}$&65$^{+60}_{-20}$&90$^{+140}_{-35}$&75$^{+70}_{-30}$&120$^{+300}_{-50}$&70$^{+21}_{-14}$&keV\\
		N$_{\rm Xill}$&9.3$\pm$2.1&7.3$\pm$2.1&8.0$\pm$1.5&14.2$\pm$2.5&12.1$\pm$2.4&14.2 $\pm$ 2.6&&$\times$10$^{-5}$ ph. cm$^{-2}$
		s$^{-1}$\\
	\end{tabular}
	
	\end{table*}
\begin{figure}
    \includegraphics[width=0.47\textwidth]{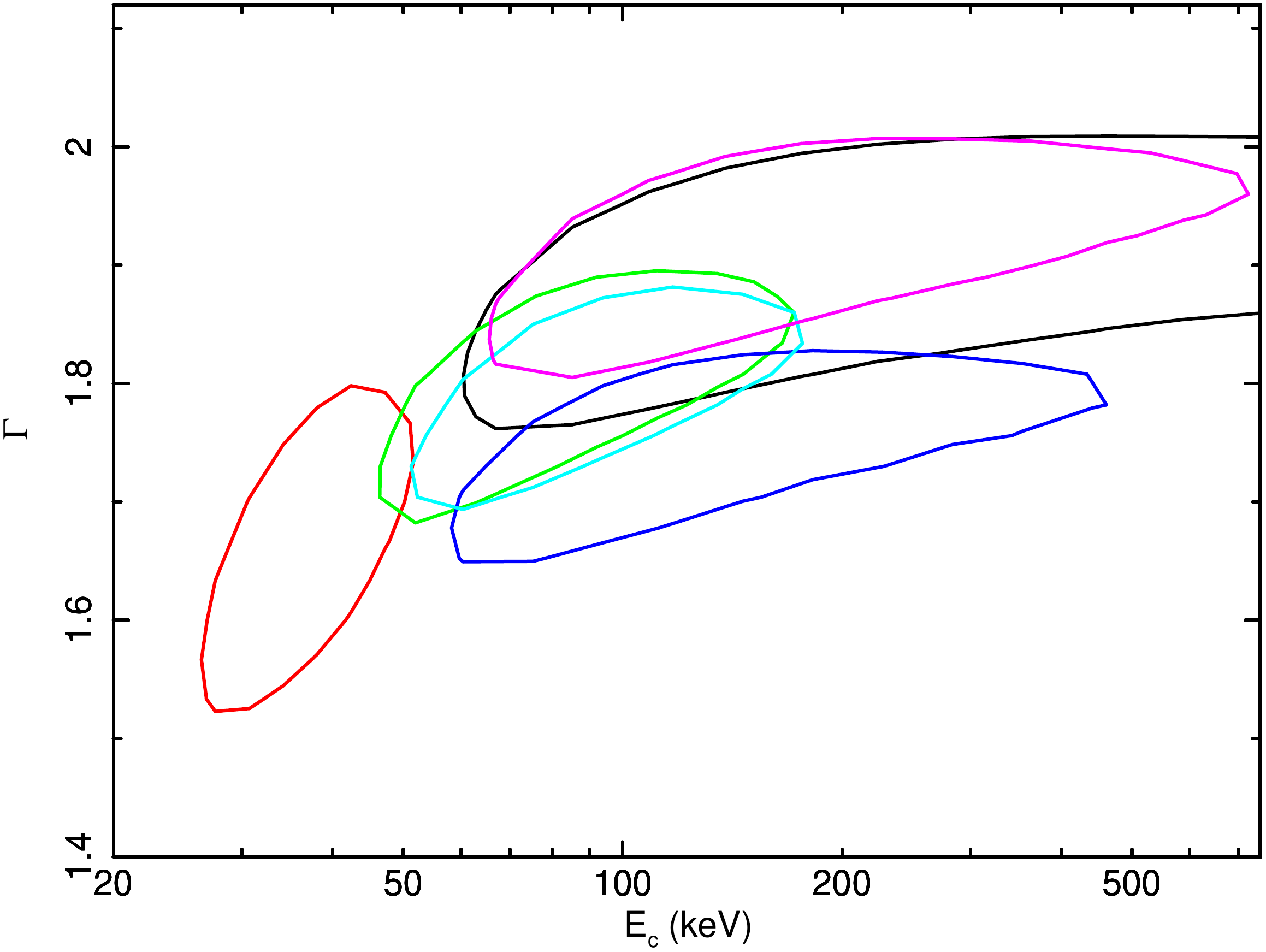}
	\includegraphics[width=0.47\textwidth]{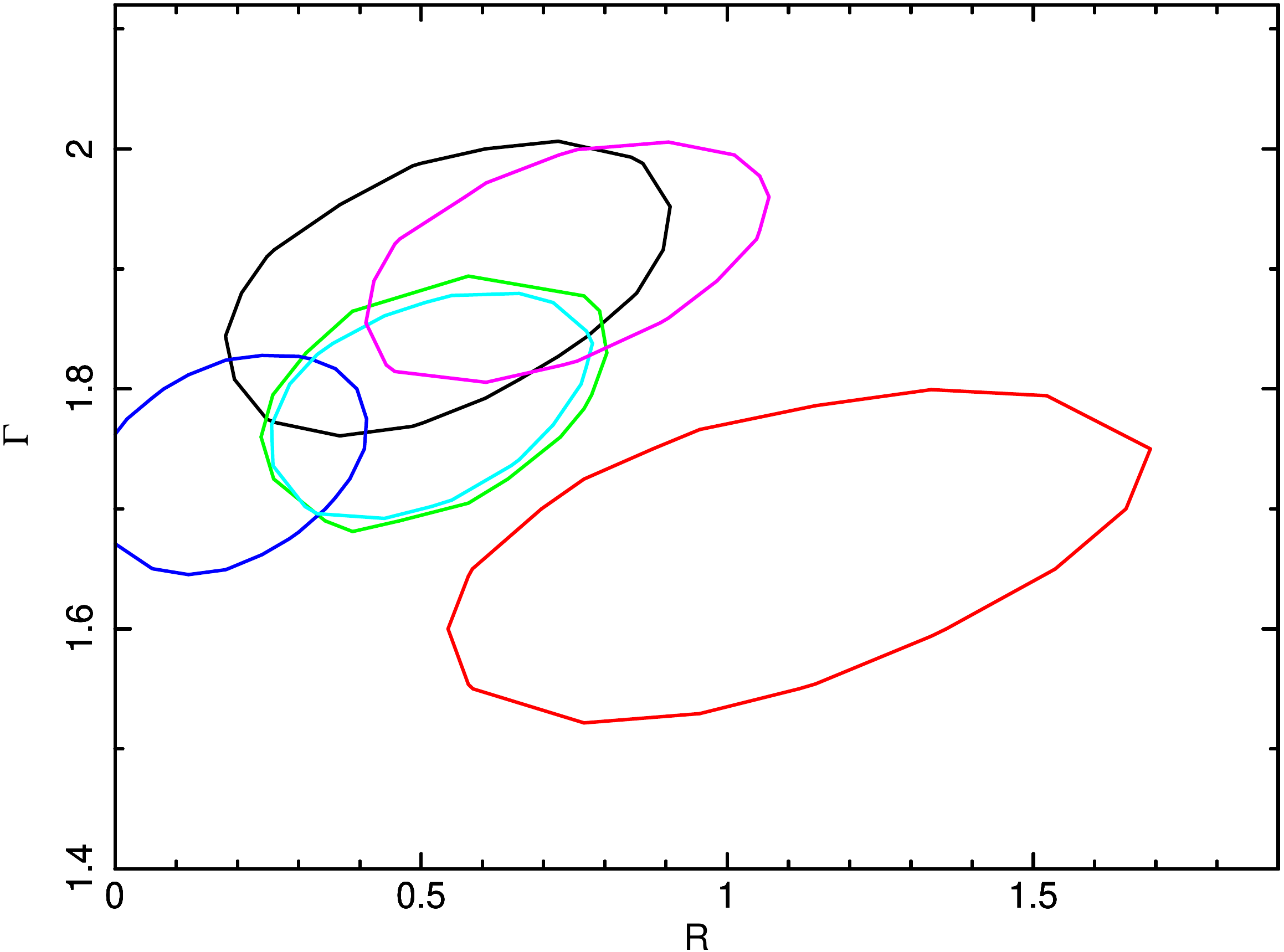}
	\includegraphics[width=0.47\textwidth]{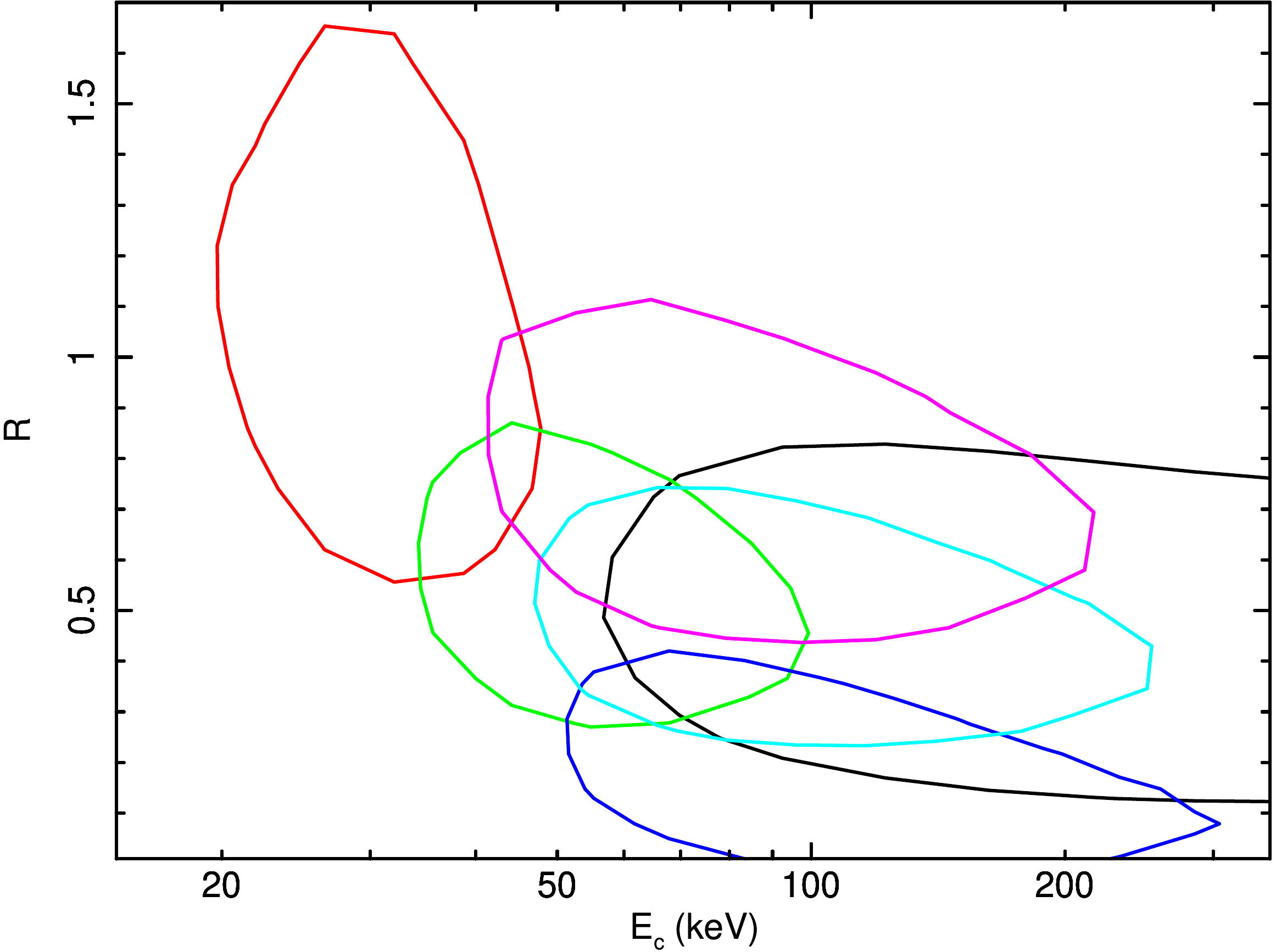}
	\caption{\small{Contours at 90\% confidence level ($\Delta\chi^2$=4.61 for two parameters) between the photon index $\Gamma$ and the high energy cut-off (E$_{\rm c}$, top hand panel) and the reflection fraction (middle). Contours in the bottom panel refer to the high energy cut-off and the reflection fraction. All the contours have been computed with the column density, the photon index, the high energy cut-off the reflection fraction and the \textit{xillver}'s normalisation free to vary in all the observations. \label{contnu}}}
\end{figure}
\subsection{Broad-band Swift/NuStar view: III the reflection signature with \textit{MyTorus}}
\begin{figure}
	\includegraphics[width=0.49\textwidth]{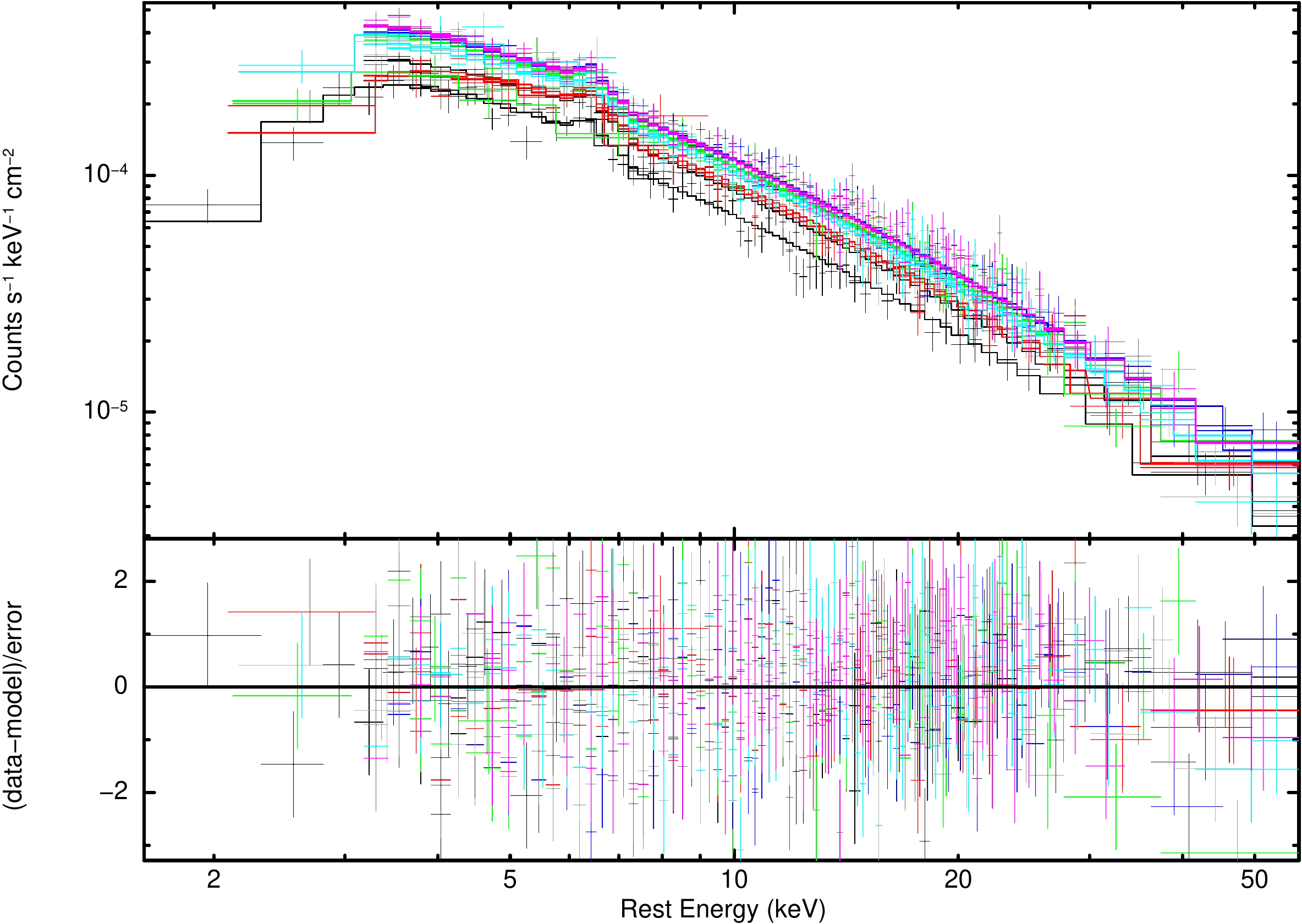}
	\caption{\small{Simultaneous \textit{Swift-NuSTAR} observations as fitted using  \textit{MyTorus} in its decoupled configuration ($\chi^2$/d.o.f.=$\chi^2$=1274 for 1258 d.o.f.).}} \label{mytorus}
\end{figure}

In the above fitting, in Sect.\,\ref{subsec:The XRT-FPMA/B 0.5-79 band I: xillver}, \textit{Xillver} assumes a geometrically simple slab reflector. Therefore, as an alternative scenario, we replaced both the \textit{Xillver} and the simple neutral absorber (\textit{ztbabs}) with \textit{MyTorus} \citep{Murphy2009}. This model takes into account the physical properties of the absorbing medium such as its toroidal geometry, the Compton-down scattering effect and it includes self-calculated reflected components (continuum plus Fe\,K emission lines). The overall \textit{MyTorus} model adopted here is composed by three publicly available tables (two additive and one multiplicative), developed for \textit{XSPEC}, which self-consistently compute the reflected continuum (\textit{MyTorusS}), the
Fe\,K$\alpha$, Fe\,K$\beta$ fluorescent emission lines (\textit{MyTorusL}) and the zeroth-order line-of-sight attenuation (\textit{MyTorusZ}). The model assumes a fixed geometry of the toroidal X-ray reprocessor, a
single value for the covering factor of the torus corresponding to an half–opening angle of 60\degree.\\
\indent We first constructed the \textit{MyTorus} model according to the `coupled’ solution \citep{Yaqoob2012}, where our line-of-sight angle that intercepts the torus is the directly co-joined (or coupled) to the scattered one. In other words, it is assumed that the fluorescent and reflected emissions emerge from the same medium that is also responsible for the line-of-sight attenuation of the X-ray underlying continuum. Thus we set the column density of each \textit{MyTorusS}, \textit{MyTorusL} and \textit{MyTorusZ} grid to be the same within each observation but free to vary between the six pointings; whilst their normalisations are tied with the one of a power-law accounting for the primary continuum. In a similar fashion, the photon-index of the three grids was linked with the nuclear one that has been computed for each observations.  These steps lead to fit to the data of $\chi^{2}$/d.o.f.=1316/1255 and considerably improves once the viewing inclination angle ($\theta_{\rm obs}$ - kept constant between the observations) is measured; by ($\Delta\chi^{2}$/$\Delta$d.o.f.$=-25/+1$). However, the returned value of line-of-sight inclination angle is $\theta_{\rm obs}=61.1_{-0.2}^{+0.4}$\,deg, just at the extremity of \textit{MyTorus}'s parameter space (i.e., $\theta_{\rm obs}=60$\,deg).\\
\indent  We then considered a more complex morphology (i.e. clumpy medium), \textit{MyTorus} further allows us to test a scenario in which the reflected and the transmitted components emerge from matter with different column densities \citep[decoupled solution see][for more details]{Yaqoob2012} in a system characterized by a more `patchy' distribution of reprocessing clumps. In fact, in such a configuration, our line-of-sight might intercept the transmitted (or zeroth-order) component through one region of the torus and the reflected emission that is back scattered from a different location of the torus itself \citep[ see][Fig. 2]{Yaqoob2012}. Such a solution, was obtained by decoupling the inclination parameter ($\theta_{\rm obs}$) of the zeroth-order (i.e., line-of-sight) and reflected table components with respective column densities defined as; $N_{\rm H,Z}$ (line-of-sight $N_{\rm H}$ ) and $N_{\rm H,S}$ (global $N_{\rm H}$).
The corresponding inclination parameters are fixed at $\theta_{\rm obs,Z}=90$\degree and $\theta_{\rm obs,S}=0$\degree for the zeroth-order and reflected continua components, respectively. In this scenario we find that the global column density is rather larger than the zeroth-order, measured at $N_{\rm H,S}=1.3_{-0.3}^{+0.5}\times10^{24}\,\rm cm^{-2}$ and the corresponding data best-fit is shown in Fig.~\ref{mytorus}. Such a model is in accordance with an emission spectrum nearly Compton-thick out the line-of-sight, and Compton-thin along the line-of-sight. The decoupled configuration yielded a $\chi^2$=1274 for 1258 d.o.f.; see Table\,\ref{torus} for the corresponding \textit{MyTorus}  best-fitted values. 
This result also suggests that the overall column density of the torus might be inhomogeneous in nature and indeed the upper parts are less dense than the central one. 
 \begin{table*}
	\centering
	\setlength{\tabcolsep}{2.0pt}
	\caption{\small{\textit{Swift-NuSTAR} best-fitted values for the parameters obtained using the decoupled \textit{MyTorus}'s solutions. The symbol $\dagger$ is used when a parameter has been fitted but tied between the observations.
	}\label{torus}}
	\begin{tabular}{c c c c c c c c c c c}
        \hline
	    Decoupled&&Parameter&Obs. 1 &Obs. 2 &Obs. 3 &Obs. 4&Obs. 5&Obs. 6 &Units\\
        &power-law&$\Gamma$&1.98$\pm$0.06&1.99$\pm$0.07&1.90$\pm$0.05 &1.90$\pm$0.06&1.94$\pm$0.05&1.89$\pm$0.06 & &\\
        %&&Norm$_{\rm scatter}$&1.1$\pm$0.8&  & &&& &$\times10^{-5}$ ph. keV$^{-1}$ cm$^{-2}$        s$^{-1}$&\\
        &\textit{MyTorusS}&N$_{\rm H}\dagger$ &1.3$_{-0.3}^{+0.5}$&&&&&&$\times10^{24}$ cm$^{-2}$ \\
        &&Norm &6.7 $\pm$1.1&8.9$\pm$2.2&6.0$\pm$1.2&9.7$\pm$1.4&10.2$\pm$2.0&9.5$\pm$1.3&$\times10^{-3}$ ph. keV$^{-1}$ cm$^{-2}$
        s$^{-1}$\\
        &\textit{MyTorusZ}&N$_{\rm H}$ &8.5$\pm$1.2&10.1$\pm$1.6 &7.0$\pm$ 1.1&7.1$\pm$1.2&8.1$\pm$1.1&6.4$\pm$1.1 & $\times10^{22}$ cm$^{-2}$ \\
	\end{tabular}
\end{table*}

\subsection{Joint \textit{XMM-Newton}, \textit{Swift} and \textit{NuSTAR} data}
\begin{figure*}
\includegraphics[width=0.49\textwidth]{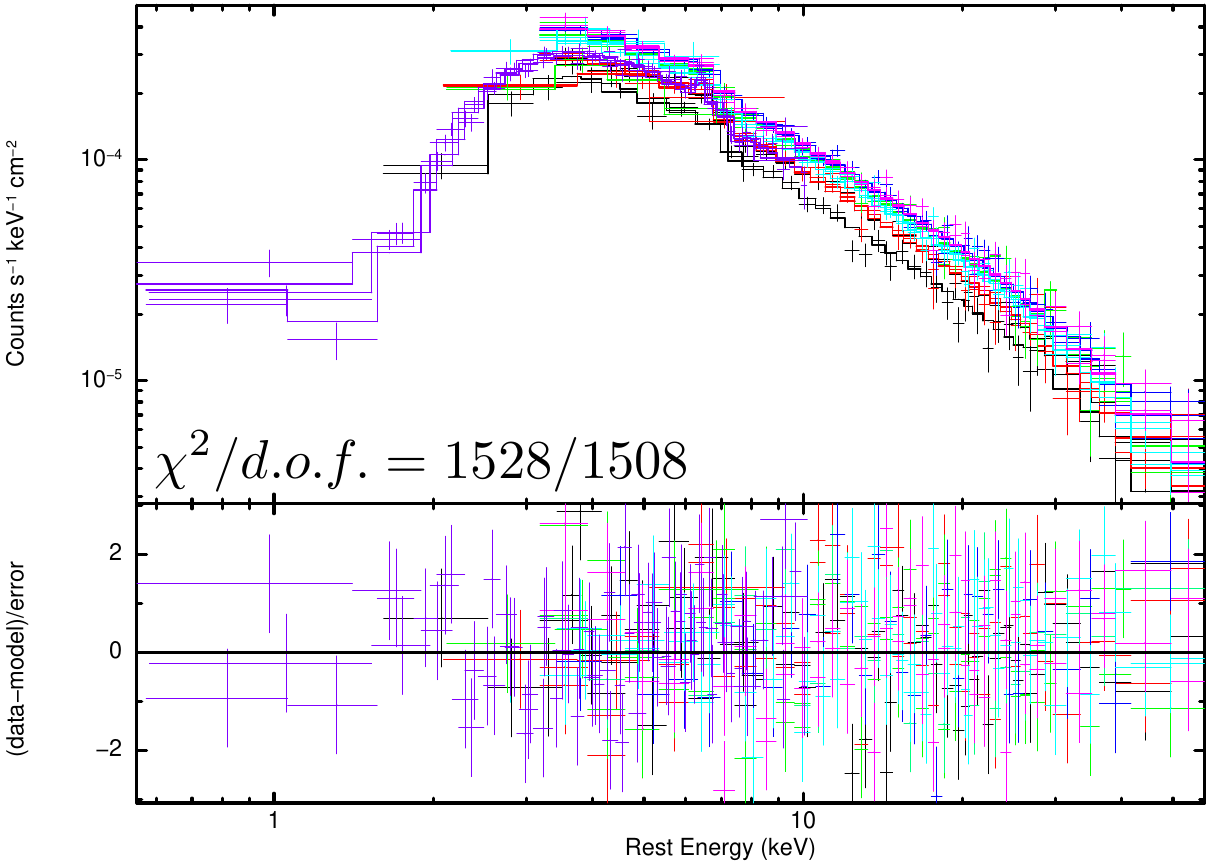}
\includegraphics[width=0.49\textwidth]{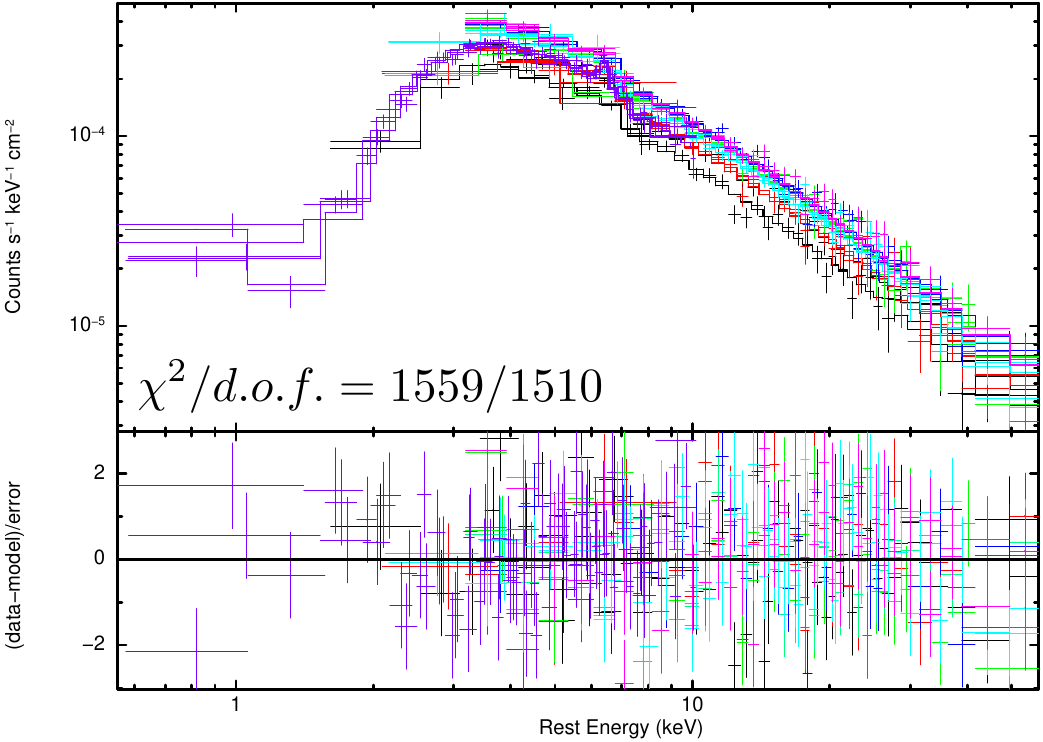}
\includegraphics[width=0.48\textwidth]{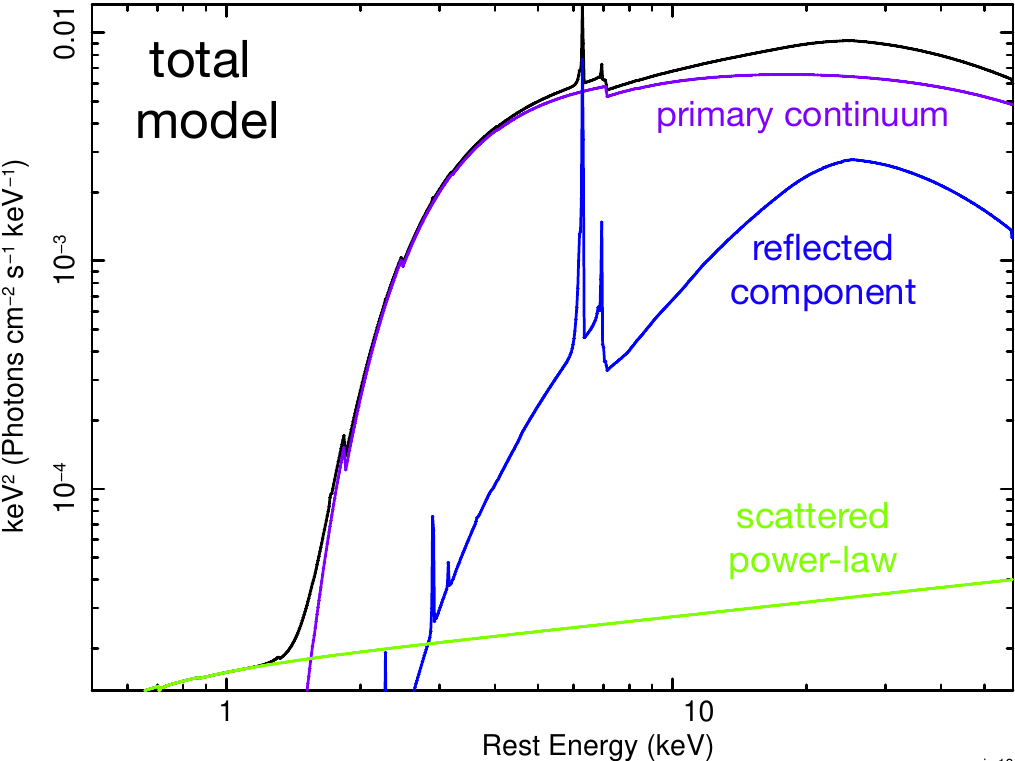}
\includegraphics[width=0.48\textwidth]{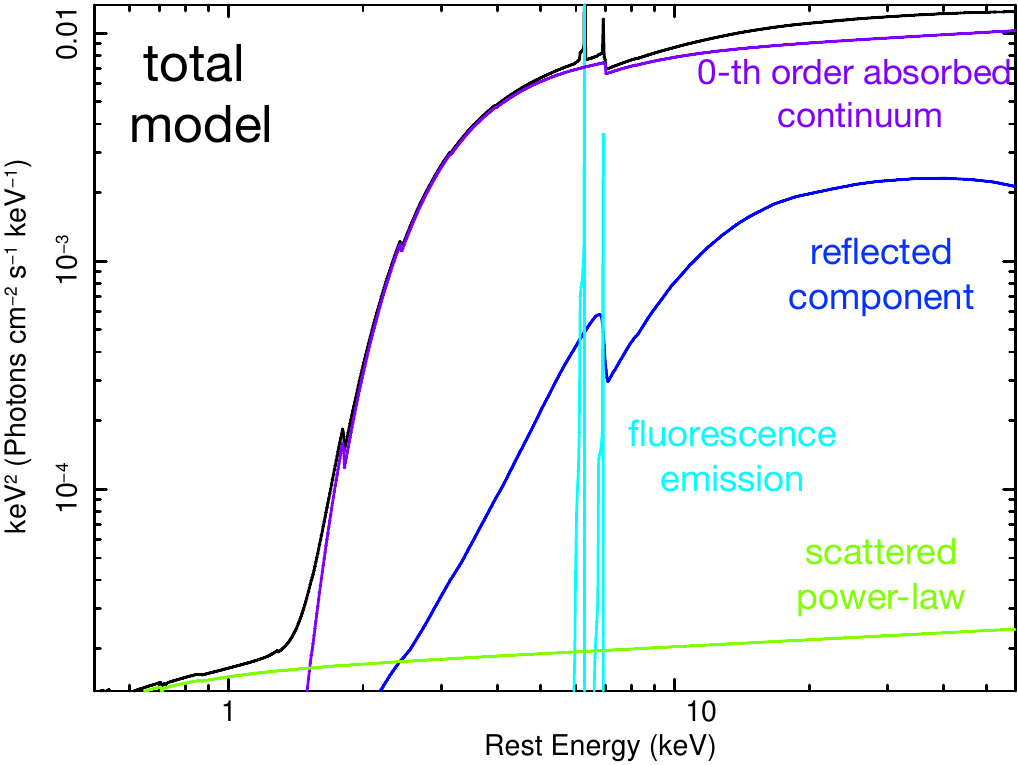}
\caption{\small{Left Panels: \textit{XMM-Newon/Swift/NuSTAR} data best-fit using \textit{xillver} (top). The different spectral components are reported in the corresponding bottom graph. Right panels: \textit{XMM-Newon/Swift/NuSTAR} data best-fit (top) and model components (bottom) corresponding to the decoupled \textit{MyTorus} solution. \label{panel}}}
\end{figure*}
The lack of substantial spectral variability between the \textit{XMM-Newton} pointing and the 2013 monitoring campaign encouraged us to perform a broadband fit based on all these data.
Therefore, we simultaneously tested the two physically motivated models in Sect. 3.3 and Sect. 3.4 on the entire data-set, i.e. on the  \textit{XMM-Newton}, \textit{Swift} and \textit{NuSTAR} spectra.\\
\indent We started using \textit{xillver} for which we fitted the photon index, the high energy cut-off and the reflection fraction tied between the pointings. We also included the scattered power-law component for which we only fitted the normalisation as its photon index was linked to the \textit{xillver}'s one. To account for the flux variability we used a constant free to vary in all but the \textit{XMM-Newton} exposure in which the \textit{xillver}'s normalisation was computed instead.
This procedure returned a fit with statistic $\chi^2$= 1528 for 1508 degree of freedom. 
The best-fit parameters were consistent with what previously found: N$_{\rm H}$=(6.6$\pm$0.2)$\times$10$^{22}$ cm$^{-2}$, $\Gamma$=1.75$\pm$0.05, E$_{\rm c}$=70$\pm$15 $\pm$ keV, A$_{\rm Fe}$=1.5$^{+1.0}_{-0.6}$, R=0.45$\pm$0.15, and the constant varied in the range 1.04-1.95.\\
\indent Then, we tested the decoupled solution of \textit{MyTorus}. We fitted the data similarly to what reported in previous Sect. 3.4 and we used a free to vary constant to account for the variations in the different observations. As we have done for the case of \textit{xillver}, we also added a scattered power-law.
These steps yielded a $\chi^2$= 1559 for 1510 d.o.f. for which the derived best fit quantities are consistent with those quoted in Table~\ref{torus}.\\
 These tests further points towards a fairly constant behaviour of the primary continuum shape and the absorbed component in MCG-01-24-12, at least for the data from 2006 to 2013. 
Although the ionised reflection model is somewhat preferred in terms of the fit statistic, both the \textit{xillver} and \textit{MyTorus} solutions give good fits. In Fig.~\ref{panel}, we report the data best-fit and the corresponding underlying model for the cases of \textit{xillver} and \textit{MyTours.}.\\
\indent As a final test, we modeled the absorption trough at $\sim$7.4 keV on the complete dataset. In particular, we used an \textit{XSTAR} \citep[][]{Kallman2004} table assuming an input spectrum of $\Gamma$=2 across the  0.1-10${^6}$ eV band and a high energy cut-off at E$_{\rm c}$=100 keV. The elemental abundance was set to the Solar one \citep{Asplund2009} and we assumed a velocity broadening v$_{\rm turb}$=5000 km\,s$^{-1}$ and the absorber to be fully covering.
In the fit, we allowed the absorber's column density, ionisation state and redshift to vary and we tied these parameters between the different observations. This additional component improved the fit with $\rm \Delta\chi^2/\Delta d.o.f.$=-11/-3, with N$_{\rm H}$=(2.3$^{+4.2}_{-1.4}$)$\times$10$^{22}$ cm$^{-2}$, $\rm \log (\xi/erg ~cm^{-2} s^{-1})$>3.1 and z$_{\rm obs}=-0.075\pm0.030$ (corresponding to a velocity v$_{xstar}$=-0.097$\pm$0.032). However, as suggested in Fig.~\ref{scannu} such an absorption feature seemed to be more prominent in Obs. 6. For this reason, we untied and fitted the absorbers parameters in this observation finding an additional enhancement of the fit statistics ($\rm \Delta\chi^2/\Delta d.o.f.$=-13/-3). The derived best fit values are consistent with each other as we obtained  N$_{\rm H}$=(1.3$_{-0.8}^{+3.2}$)$\times$10$^{23}$ cm$^{-2}$, $\rm \log (\xi/erg ~cm^{-2} s^{-1})$=3.2$\pm$0.4 for z$_{\rm obs}$=-0.076$\pm$0.018. By untying these parameters across all the observations, would lead to a marginal improvement of $\rm \Delta\chi^2/\Delta d.o.f.$=-21/-15. Finally, these values are fully compatible within each other and with the parameters commonly measured for these absorbers in other AGNs \citep[e.g.][]{Gofford2013,Tombesi2013}.

\subsection{The 2019 ACIS/S spectrum}

The poor statistics of the Chandra snapshot did not allows us to do a detailed spectral analysis. In fact, a simple phenomenological model such as an absorbed power-law fails in reproducing the hard continuum and returned a photon index $\Gamma\lesssim$1.  Then, we tested a scenario in which the source had an intrinsic flux drop likely due to a change in its luminosity. To do this, we applied the best-fit model used for the EPIC data (see~\ref{xmmfit}) on the  \textit{Chandra} spectrum and we refit this data only allowing the primary normalisation to vary.
This procedure yielded a fit with $\chi^2$=56 for 29 d.o.f. and returned a primary normalisation N$_{\rm po}$=(7.0$\pm$0.5)$\times10^{-4}$ ph. keV$^{-1}$ cm$^{-2}$ s$^{-1}$ and an observed flux F$_{\rm 2-10~keV}$=(2.15$\pm$0.15)$\times10^{-12}$ erg cm$^{-2}$ s$^{-1}$, a factor of 10 lower than what previously measured.
Such an intrinsic flux drop seems to be favoured with respect to a scenario in which the N$_{\rm H}$ of the neutral obscurer changed: keeping fixed the normalisation of the primary power-law at the best-fit value as reported in Table \ref{xmmfit} and letting free to vary only the column density of the neutral obscurer returned a $\chi^2/d.o.f. >$11.
A simultaneous fit of both the primary continuum normalisation and the obscuring column led to a fit statistic of  $\chi^2$/d.o.f.=53/28 with N$_{\rm H}$=7.4$\pm$1.0 $\times$10$^{22}$ cm$^{-2}$ and  N$_{\rm po}$=(7.5$\pm$0.8)$\times10^{-4}$ ph. keV$^{-1}$ cm$^{-2}$ s$^{-1}$. The fit of the Fe K$\alpha$ energy centroid normalisation enhanced the fit by $\Delta\chi^2/\Delta$d.o.f.=-12/-2. The Fe K$\alpha$ has energy E=6.5$\pm$0.1 keV, a normalisation N$_{\rm Fe~K\alpha}$=(1.1$\pm$0.8)$\times$ 10$^{-5}$ ph. cm$^{-2}$ s$^{-1}$  and EW=320$\pm$230 eV with this feature being consistent within the errors with what previously observed. The limited bandwidth of the data did not allow us to further investigate the presence of any absorption features nor to firmly determine the physical origin of such
a low flux state.

\section{Discussion and Conclusions}

We reported on the X-ray emission properties of the Seyfert 2 galaxy MCG-01-24-12 based on observations taken with several X-ray facilities over a time interval spanning about 13 years. In the following we summarise and discuss our findings.

\subsection{Variability and phenomenological modelling}

\textit{XMM-Newton} and \textit{NuSTAR} data are consistent with a moderate variability  of the source flux in the 3-10 keV band with values in the range of 1.2-2.3$\times$10$^{-11}$ erg cm$^{-2}$ s$^{-1}$. Interestingly this flux state is consistent with what was measured using \textit{BeppoSAX} data \citep[][]{Malizia2002} and by \citet[][]{Piccinotti82}.\\
We computed the bolometric luminosity and Eddington ratio of the source assuming an average flux state which corresponds to a luminosity L$_{\rm 2-10 ~ keV}$ $\sim$1.5$\times$10$^{43}$ erg s$^{-1}$. Following the prescription in \citet{Duras2020} and using a SMBH mass M$_{\rm BH}$=1.5$\times$10$^{7}$ M$_{\sun}$ \citep{LaFranca2015}, we derived L$_{\rm Bol}\sim$1.9$\times$10$^{44}$ erg s$^{-1}$ and L$_{\rm Bol}/L_{Edd}\sim$11\%, respectively. 
As displayed in Fig.~\ref{lc}, variations occurred on weekly time scales and intra-observation changes are weak, with the only exceptions of Obs. 2 and 3 where the hardness ratios also suggest some spectral change down to ksecs timescales. On the other hand, the 2019 exposure performed with \textit{Chandra} caught the source in an unprecedented observed faint state (L$_{\rm 2-10 ~ keV}$ $\sim$2$\times$10$^{42}$ erg s$^{-1}$ ). The source faded by a factor of $\sim$10 from the last \textit{NuSTAR} exposure. Different  physical origins explained remarkable variations in other AGNs: (i) an increase in the neutral obscuration in which the column density swings from Compton-thin to thick in timescales of hours-days as seen in the prototype changing-look AGN NGC\,1365 \citep[see][]{Risaliti05b}, (ii) an obscuration event due to the clumpy highly ionised disc-wind as seen in MCG-03-58-007 \citep{Braito2018,Matzeu19}, (iii) strong intrinsic variability but neutral N$_{\rm H}$ fairly constant see e.g. the case of NGC\,2992 discussed in \citet{Murphy2007} and Middei et al., in preparation.\\
\indent  From a phenomenological prospective, the primary emission of MCG-01-24-12 is consistent with an absorbed power-law where the column density and spectral shape had a fairly constant behaviour within the \textit{NuSTAR} monitoring and when these 2013 data are compared with those from \textit{XMM-Newton} and \textit{BeppoSAX}. The Fe K$\alpha$ emission line seems to vary in the \textit{NuSTAR} spectrum and is strongly detected in Obs. 2 and 6. In this respect, we notice that the strongest Fe\,K$\alpha$ is observed in Obs. 2 (see Table~\ref{pippo} and Fig.\ref{scannu}) and that the line flux seems not to follow the weak variations of the continuum. Such a behaviour can be explained by the reverberation of the Fe K$\alpha$ line \citep[see e.g.][]{Zoghbi2019} that, produced in a distant region such as the BLR, would have a delayed response with respect to the primary continuum.\\
\indent The short duration of the exposures coupled with the instrumental spectral resolution prevented a detailed analysis on line profile that was set to be narrow. On the other hand, the EPIC data are consistent with a moderately broad Fe K$\alpha$ emission line ($\sigma$=80$\pm$60 eV) corresponding to a region of some hundredths of a parsec\footnote{This approximate estimate is derived via the Virial Theorem from which the Fe K$\alpha$ emission line is originated at r=(E/$\rm \Delta E)^2 \rm \times r_g$. In our case, this estimate returns r$\sim$1.4$\times$10$^{16}$ cm.}. Interestingly, troughs above 7 keV have been observed in both \textit{NuSTAR} and \textit{XMM-Newton} spectra, possibly suggesting for the presence of fast and highly ionised outflows.\\

\subsection{Physically motivated modelling}
\indent \textit{Xillver} provided the best fit to the \textit{Swift}-\textit{NuSTAR} data, i.e. a cut-off power law continuum plus its associated Compton reflected spectrum absorbed by a column density of about ($6.3\pm0.5$)$\times10^{22}$\,cm$^{-2}$. The chemical abundance of the reflecting material is consistent with being Solar (A$_{\rm Fe}$=1.3$\pm$0.6) and the continuum photon index, high energy cut-off and reflection fraction are constant within the errors in all but Obs. 2, see Fig.~\ref{contnu}. This observation is the only one characterised by a variable ratio of the 3-10 and 10-79 keV light curves and this possibly explains the discrepancies between Obs. 2 and the other observations. Such a case of short term spectral variability is quite peculiar for MCG-01-24-12 since the analysis of the other exposures agrees with an intra-observations constancy of the primary photon index.
However, when the fit is performed with the photon index, the high energy cut-off and the reflection fraction tied between the observations a statistic of $\chi^2$= 1300 for 1278 d.o.f., with the best-fit obtained computing the various parameters in all the observations.\\
\indent For this reason, we used the average values for the photon index ($\Gamma$=1.76$\pm$0.09) and the high energy cut-off (E$_{\rm c}=70^{+21}_{+14}$ keV \citep[these values are also consistent with those found in][]{Balokovi2020} to derive the properties of the hot corona. The physical conditions of the emitting plasma are indeed responsible for the spectral shape and the high energy curvature of the X-ray continuum \citep[e.g.][]{Rybi79,Belo99,Petr00,Petr01,Ghisellini13,Middei2019b}. The relations between $\Gamma$-E$_{\rm c}$ and kT$_{\rm e}$-$\tau_{\rm e}$ have been recently derived by \citet{Middei2019b} that used extensive simulations computed with \textit{MoCA} \citep[Monte Carlo code for Comptonisation in Astrophysics,][]{Tamborra2018} for studying the Comptonised spectrum of AGNs \citep[see also][for further applications of this code]{Marinucci2019,Lanzuisi2019}. In particular, using Eqs. 2, 3, 4 and 5 in \citet{Middei2019b}, we found the hot corona in MCG-01-24-12 to be characterised by kT$_{\rm e}$=27$^{+8}_{-4}$ keV, $\tau_{\rm e}$=5.5$\pm$1.3 and kT$_{\rm e}$=28$^{+7}_{-5}$ keV, $\tau$=3.2$\pm$0.8 for a spherical emitting plasma and a slab-like one, respectively. These values are fully in agreement with average temperature and opacity found in literature \citep[e.g.][]{Fabi15,Fabi17,Tortosa18,Middei2019b}. Then, we used the coronal temperature and opacity to include this source in the compactness-temperature l-$\theta_{\rm e}$) diagram \citep[e.g.][]{Fabi15,Fabi17}. We calculate the dimensionless coronal temperature $\theta_{\rm e}$=kT$_{\rm e}$/m$_{\rm e}$c$^2$ and the compactness parameter $l$ = L$\sigma_{\rm T}$/Rm$_{\rm e}$c$^3$, in which L accounts for the coronal luminosity in the 0.1-200 keV band and R for its radius that is assumed to be 10 gravitational radii (R10). Following these prescriptions, we derived $\theta_{\rm e}$=0.053$^{+0.015}_{-0.08}$ and $l\simeq$55. These values are in agreement with the bulk of measurements presented by \cite{Fabi15} and show that the source lies in the so-called permitted region in which annihilation is still dominant with respect to the pair production.\\
\indent We find \textit{MyTorus} can provide a good representation of the \textit{Swift-NuSTAR} data. In fact, although the coupled solution is not adequate to described the overall spectrum of MCG-01-24-12 (with this mainly due to the relatively small curvature at low energies of \textit{NuSTAR} data and the poor statistic of \textit{XRT} spectra), the decoupled configuration provides a statistically acceptable representation of the \textit{Swift-NuSTAR} spectra of MCG-01-24-12. We measured a considerable difference between the column densities of the global (out of the line-of-sight) and transmitted reprocessors. The nuclear radiation is absorbed by a neutral medium with the column density N$_{\rm H,Z}$ in the range (5.3-11.7)$\times$10$^{22}$\,cm$^{-2}$ and the reflected component is back mirrored by matter with global N$_{\rm H,S}=$1.3$^{+0.5}_{-0.3}\times10^{24}$\,cm$^{-2}$. Such distinction between the zeroth-order and the global density has been already observed in other Seyfert\,2 galaxies, e.g. NGC\,4945, \citep[][]{Yaqoob2012}, Mrk\,3 \citep[][]{Yaqoob15}, MCG-03-58-007, \citep[][]{Matzeu19}, NGC 4507 (Zaino et al., sub.), NGC\,4347 \citep[][]{Kammoun19} and described further in \citet{Kammoun20arxiv}. Thus the emerging picture is consistent with an overall inhomogeneous or `patchy' toroidal absorber, broadly Compton-thin, with a distribution of relatively small and thicker equatorial clouds out of the line-of-sight. In most recent torus models the `viewing probability' of the absorber, which is strongly correlated on its size and location, tend to be typically distributed towards the equatorial plane. The inhomogeneous gas distribution of the torus is now a well established scenario within the scientific community, where a variety of models have been developed taking into account this physical framework \citep[e.g., ][]{Elitzur06,Nenkova08a,Nenkova08b,Tanimoto19}.

\subsection{Is there a variable wind in MCG-01-24-12?}
\indent The inhomogeneous nature of the absorber and a viewing angle that possibly just passes through a semi-transparent obscurer, allowed us to observe the nuclear regions of the MCG-01-24-12. Such a framework was found suitable for the star-forming galaxy MCG-03-58-007 \citep{Braito2018}. This source hosts a very powerful ($v_{\rm out}\sim0.1-0.3$~c), variable ($\delta t\lesssim \rm 1 day$) and multi-structured disc-wind launched between tens to hundreds of gravitational radii from the black hole \citep[see][Fig. 9]{Matzeu19}. The possible detection of blue-shifted absorption lines on \textit{XMM-Newton} and \textit{NuSTAR}\,Obs\,6 spectra may suggest MCG-01-24-12 to be similar to MCG-03-58-007. If the absorption troughs are associated with Fe\,\textsc{XXVI} Ly$\alpha$ they would correspond  to a highly ionised gas outflowing at $v_{\rm out}\sim$($0.06\pm0.01$)c \footnote{This outflowing velocity has been derived in accordance with $v_{\rm z_{abs}}= \left( (1+z_{abs})^{2} - 1\right)/\left( 1+z_{abs})^{2} + 1\right)$ and
$v_{\rm out}/c = \left(u - v_{\rm z_{abs}}\right)/\left(1-(uv_{\rm z_{abs}})\right)$; where z$_{abs}$ is the redshift of the feature and $u$ is the systemic velocity of MCG-01-24-12.}.
Interestingly, in accordance with the blind line scan, the absorption feature above $\sim$7 keV has a significance $\gtrsim$90\% in 3 \textit{NuSTAR} observations and in the \textit{XMM-Newton} one. These troughts, together with the one found by \cite{Malizia2002} in \textit{BeppoSAX/PDS} data, may suggest for the presence of a persistent wind.
We finally noticed that a powerful disc-wind has been invoked to explain a low flux state observed in MCG-03-58-007  \citep[][]{Braito2018,Matzeu19} where the authors found the source variability to be caused by a highly ionised fast wind rather than by a neutral clumpy medium. The low counts of the \textit{Chandra} snapshot did not allow us to test such a scenario and a longer \textit{XMM-Newton} exposure is needed  to confirm the putative outflow in MCG-01-24-12 and to further understand the physics behind its low flux state. Moreover, future observations through the high resolution micro-calorimeter detectors on board of \textit{XRISM} and \textit{ATHENA} will provide unprecedented details of these obscuration events.     

\begin{acknowledgements}
We thank the anonymous referee for useful comments.
RM thanks Fausto Vagnetti for discussions and insights and Francesco Saturni for useful comments. RM acknowledges the financial support of INAF (Istituto Nazionale di Astrofisica), Osservatorio Astronomico di Roma, ASI (Agenzia Spaziale Italiana) under contract to INAF: ASI 2014-049-R.0 dedicated to SSDC. Part of this work is based on archival data, software or online services provided by the Space Science Data Center - ASI. S.B. acknowledges financial support from ASI under grants ASI-INAF I/037/12/0 and ASI-INAF n.2017-14-H. A.D.R.  acknowledges financial contribution from the agreement ASI-INAF n.2017-14-H.O. SB, ADR MD, AM and AZ acknowledge support from PRIN MIUR project "Black Hole winds and the Baryon Life Cycle of Galaxies: the stone-guest at the galaxy evolution supper", contract no. 2017PH3WAT.  Part of this work is based on archival data, software or online services provided by the Space Science Data Center - ASI.  This  work  is  based  on  observations obtained with: the NuSTAR mission,  a  project  led  by  the  California  Institute  of  Technology,  managed  by  the  Jet  Propulsion  Laboratory  and  funded  by  NASA; XMM-Newton,  an  ESA  science  mission  with  instruments  and  contributions  directly funded  by  ESA  Member  States  and  the  USA  (NASA).

\end{acknowledgements}

\thispagestyle{empty}
\bibliographystyle{aa}
\bibliography{MCG_1_24_12.bib}

\end{document}